\documentclass{article}
\usepackage[utf8]{inputenc}

\usepackage[margin=2cm]{geometry}
\usepackage[utf8]{inputenc}
\usepackage{url}
\usepackage{array}
\usepackage{amsmath,amssymb,amsfonts}
\usepackage{mathtools}
\usepackage{algorithm}
\usepackage{algpseudocode}
\usepackage{tikz}
\usepackage{graphicx}
\usepackage{subcaption}
\usepackage[labelfont=bf]{caption}
\usepackage[export]{adjustbox}
\usepackage{bm}
\usepackage{xcolor}
\usepackage{tabularx}
\usepackage{hyperref}
\usepackage{pdflscape}
\usepackage{afterpage}
\usepackage{authblk}
\usepackage[symbol]{footmisc}
\usepackage{comment}
\usepackage{array}
\newcolumntype{H}{>{\setbox0=\hbox\bgroup}c<{\egroup}@{}}
\usepackage[parfill]{parskip}
\usepackage[backend=bibtex,style=science]{biblatex}
\def\dt{\Delta t}
\def\MSD{\mathrm{MSD}}
\def\VACF{\mathrm{VACF}}

\title{\textbf{Anomalous Platelet Transport \& Fat-Tailed Distributions: A Paradigm Shift}}

\author[1]{C. Kotsalos\thanks{Corresponding author: kotsaloscv@gmail.com}}
\author[2]{K. Z. Boudjeltia}
\author[3]{R. Dutta}
\author[1]{J. Latt}
\author[1]{B. Chopard}

\affil[1]{\small University of Geneva, Computer Science Department, Scientific and Parallel Computing Group.}
\affil[2]{\small Laboratory of Experimental Medicine (ULB222), Faculty of Medicine, Université libre de Bruxelles, CHU de Charleroi.}
\affil[3]{\small University of Warwick, Department of Statistics.}

\date{\vspace{-5ex}}

\begin{document}

\maketitle

\noindent\makebox[\linewidth]{\rule{\textwidth}{0.5pt}}
\textbf{Abstract}

The transport of platelets in blood is commonly
assumed to obey an advection-diffusion equation. Here we propose a
disruptive view, by showing that the random part of their velocity is
governed by a fat-tailed probability distribution, usually referred to
as a Lévy flight. Although for small spatio-temporal scales, it is
hard to distinguish it from the generally accepted ``red blood cell enhanced''
Brownian motion, for larger systems this effect is dramatic as the
standard approach may underestimate the flux of platelets by several
orders of magnitude, compromising in particular the validity of
current platelet function tests.

\noindent\makebox[\linewidth]{\rule{\textwidth}{0.5pt}}

\section{Introduction}
Platelets are entities involved in multiple physiological and pathophysiological processes such as haemostasis, thrombosis, clot retraction, vessel constriction and repair, inflammation including promotion of atherosclerosis, host defense, and even tumor growth/metastasis \cite{Harrison2005}. Platelets (PLTs) are the second most numerous cell in blood, after red
blood cells (RBCs), with concentration of 150-450$\times 10^9 /l$. Their 
size, shape, material and transport properties allow them to be optimally placed as close as
possible to the vessel wall, a physical requirement for the constant
inspection of the integrity of the vasculature. Upon injury platelets respond
rapidly (through activation, adhesion, aggregation, release reactions,
etc.), and form a haemostatic plug, occluding the damaged site and
preventing blood loss. Any disorder in these physiological processes
results in impaired haemostasis, and inappropriate thrombus
formation. For example, arterial thrombi can develop within
atherosclerotic lesions resulting in stroke and heart attack, two of
the major causes of morbidity and mortality in the western world
\cite{Harrison2005}.

In view of the involvement of platelets in the formation of thrombi, it is necessary  to develop antiplatelet drugs. These drugs have been used in primary and secondary prevention for several decades now. Despite this, it appears that some patients under treatment still have a cardiovascular event. It has been therefore sought to assess the effect of these treatments and their dosage by means of platelet function tests. The results were not very conclusive. Indeed, a review article \cite{Breet2010} has evaluated platelet functions using 6 different techniques in patients undergoing coronary stent implantation. The correlation between the clinical biological measures and the occurrence of a cardiovascular event was null for 3 of the techniques, and rather modest for the 3 others, indicating the obvious need for a more efficient tool able to monitor patient's platelet functionalities. In another review \cite{Picker2011}, the author insists on the fact that no current test allows the analysis of the different stages of platelet activation or the prediction of the in-vivo behavior of those platelets. A more recent publication \cite{Koltai2017} reported that controlled trials have consistently failed to demonstrate a benefit of personalized antiplatelet therapy based on platelet function testing.

The role of PLTs to repair blood vessels thanks to their capability to adhere and aggregate on damaged tissues cannot be isolated from the presence of RBCs. In addition to delivering oxygen, RBCs have a significant  influence on blood clotting and thrombosis \cite{Weisel2019}. As a particular example of the involvement of RBCs is a rheological effect known as platelet margination. In blood flow, RBCs induce a movement of platelets from the center of blood vessel to the wall. Therefore, PLTs are adjacent to the vessel wall, where they can interact to form transitory plugs in case of injury \cite{Flamm2012}. Besides, this layer at the surface of the endothelium also contains plasma with coagulation factors and white cells.

Knowing how platelets travel in blood while colliding with other blood cells is still a crucial question for scientists. Indeed,  how does a doctor know whether the dysfunction of platelets is only due to a modification of platelets adhesion/aggregation capabilities or due to a change of their transport properties induced by pathological RBCs? To test whether platelets dysfunction, one has to disentangle the adhesion/aggregation from their movement, as both processes are crucial for their proper functioning. Since current platelets tests, in clinical laboratory routine, do not take into account the role of RBCs in platelet transport and adhesion/aggregation processes, a better understanding and description of the RBCs-PLTs interactions seems critical to develop a new generation of platelet function tests closer to physiological reality.

In order to make a significant step in this direction, the focal point of this paper is to reveal the true mechanism of platelet transport. As we shall see, the current assumption that platelets are subject to an advection-diffusion process in the blood has to be revised. 

Platelets move within blood due to the combined effect of
the plasma velocity (advection) and the collisions with RBCs (enhanced diffusion). In particular, in
a shear flow, PLTs experience a random motion in the
direction perpendicular to the flow. The accepted description of this
process (the so-called Zydney-Colton theory \cite{ZydneyColton})
is that PLTs are subject to a diffusion process, whose diffusion
coefficient is \cite{Affeld2013}
\begin{equation}
  D = D_{PRP} \times (1-H) + 0.15 \times (d_{RBC}^2 \times H/4) \times \dot{\gamma} \times (1-H)^{1.8}
\label{eq:ZC}
\end{equation}
where $D_{PRP}$ is the diffusivity of PLTs in a platelet-rich plasma
(without RBCs) and its value is typically
$D_{PRP}=\mathcal{O}(10^{-13})~m^2/s$. The quantity $\dot{\gamma}$ is
the shear rate, $H$ is the hematocrit, and $d_{RBC}$ is the diameter
of a RBC.

Equation (\ref{eq:ZC}) gives $D=\mathcal{O}(10^{-11})$ $m^2/s$,
for the situation described in Chopard et al. \cite{Chopard2017},
namely the diffusion of PLTs in a shear flow with
$\dot{\gamma}=100~s^{-1}$ and $H=0.35$, as created by the so-called
impact-R PLT analyzer. This device allows one to measure the amount
of PLTs that deposit on a surface perpendicular to the flow.  However,
based on such experimental evidence, Chopard et al. showed that if one assumes
that the concentration $\rho$ of PLTs obeys the diffusion equation
\begin{equation}
  \partial\rho=D\nabla^2\rho
  \label{eq:diffusion}
\end{equation}
a value of $D \approx \mathcal{O}(10^{-8})$ $m^2/s$ is needed to
explain the number of platelets that is observed to deposit in impact-R
platelet analyzer.

This result obviously raises the question of the validity of
eq. (\ref{eq:ZC}) or the applicability of eq. (\ref{eq:diffusion}).
The Zydney-Colton model results from accumulating
experimental/theoretical data, and it has been extensively validated
by numerous numerical studies, in which RBCs and PLTs where resolved
\cite{Crowl2011,Zhao2011,Zhao2012,Reasor2013,Vahidkhah2014,Mehrabadi2015,Mehrabadi2016}. However,
these studies concern spatio-temporal scales much smaller than those
characterizing the impact-R device. The latter considers a layer of
blood of $820~\mu m$ of thickness, rotating in a cylinder of diameter
$6.5~mm$. The amount of platelets that disappear from the bulk due to
their deposition on the bottom part of the cylinder is observed after
$20~s$. State of the art numerical simulations consider much smaller
systems, usually less than $100~\mu m$ in size, for less than $1~s$ of
physical time.

A possibility to explain this 1000-fold difference between the
Zydney-Colton theory and the effective observed diffusion is to
postulate the presence of a drift term in addition to the diffusion
process, as suggested in \cite{Eckstein1991,Crowl2011,Kumar2012}, 
as a general mechanism for the case of transport of deformable
suspensions. This drift-diffusion model, which includes a
``rheological potential'' ($\Phi$), has however an ill-considered/posed
origin.  Furthermore, due to the symmetry of the problem it is hard to
understand why such a symmetry breaking would appear in the case of
the impact-R. As a matter of fact, simulations of PLT transport in
the impact-R, including an ad hoc drift term and a Zydney-Colton
diffusion, do not fit so well the in-vitro time evolution of PLTs
deposition (data from B. Chopard 2018).

Our claim in the present paper is that it is possible to reconcile
these contradictory results by assuming that PLTs do not follow a
Gaussian random walk (as is implied by the standard diffusion
equation), but a random walk with a fat-tailed distribution of
velocities.  We show here that a very careful analysis of fully
resolved blood flow simulations, in which deformable RBCs and platelets
interact and move in a suspending fluid (the plasma) reveals that
collisions between RBCs and PLTs result in a power law probability
distribution of PLT velocities,
\begin{equation}
  P(v)\sim v^{-1-\alpha}
\end{equation}
with $\alpha$ around 1.5 (power law exponent). We also show that for small systems, the value of the PLT mean square displacement (MSD), which is
traditionally related to the diffusion coefficient, is compatible with
the Zydney-Colton theory, explaining why the normal diffusion
hypothesis was little questioned in the literature. But as the system
size increases, we observed that the MSD increases, a behavior
incompatible with a Gaussian random walk.

It should be noted that, despite this wide consensus that
Zydney-Colton theory applies, a few researchers have superficially
pointed out events that support our current result. Vahidkhah et
al. \cite{Vahidkhah2014,Vahidkhah2015} observed highly
anisotropic RBC distribution and a ``waterfall'' phenomenon (cavities
that act as express lanes for platelets) affecting PLT movement
(supported experimentally by Lee et
al. \cite{Lee2013}). Mehrabadi et al. \cite{Mehrabadi2015}
pointed out a possible anomalous diffusion as PLTs get trapped in the
cell free layer. Yeo et al. \cite{Yeo2010} (same approach as
Gross et al. \cite{Gross2014}) who talked about anomalous
diffusion of wall-bounded non-colloidal suspensions, observed
exponential distributions for the densely packed spherical particles
(correspondence to RBCs).

\begin{figure}[h]
    \centering
    \includegraphics[scale=0.4]{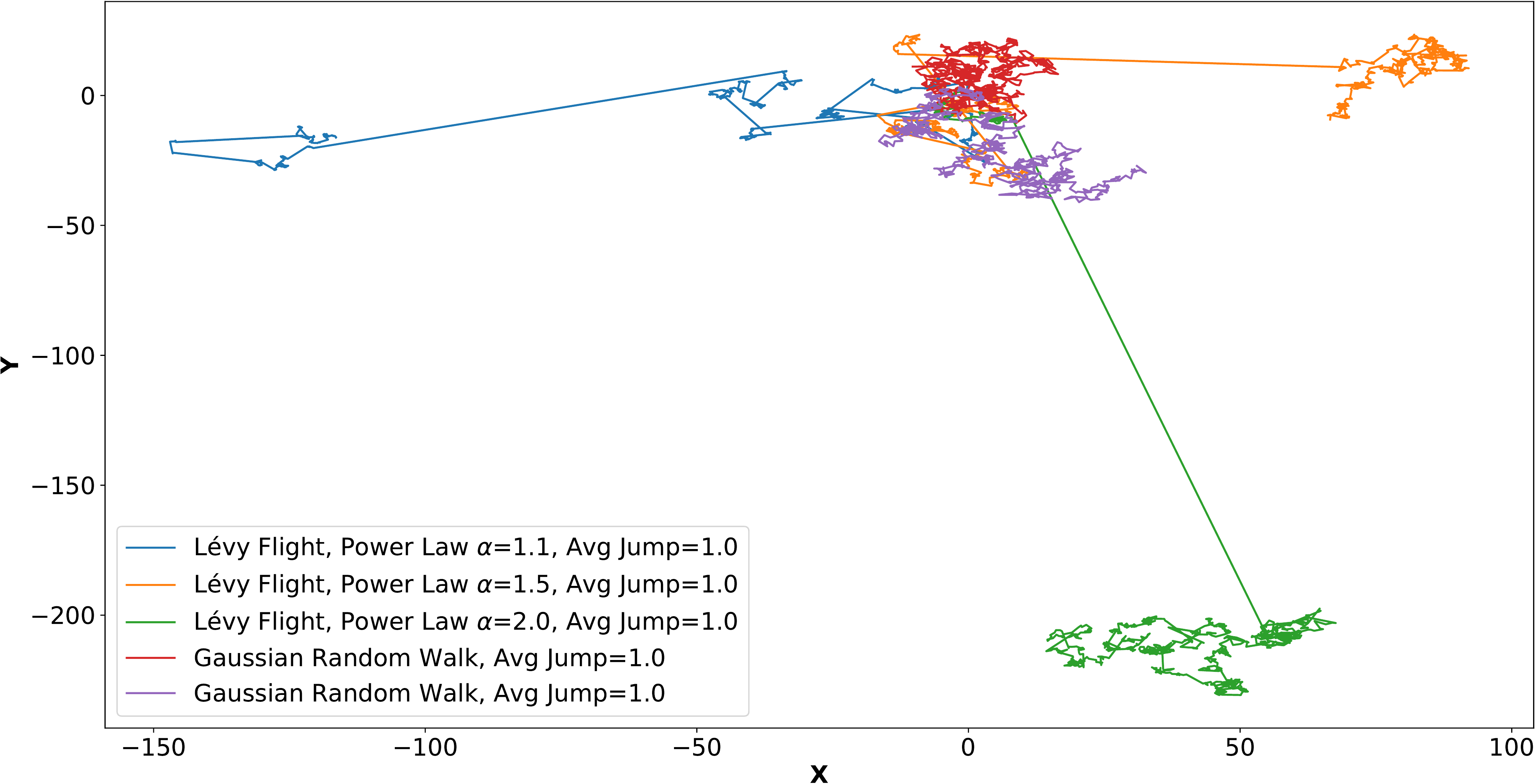}
    \caption{\textbf{Random Walks with different underlying distributions.} Five particles performing random walks, three of them exhibit power law behavior with varying exponent and two of them follow a Gaussian velocity distribution. The exploration of the space is largely dependent on the underlying probability density function.}
    \label{fig:powerlaw_walkers}
\end{figure}

It is critical to understand, why an underlying power law behavior leads to enormous differences in transport physics. In Fig. \ref{fig:powerlaw_walkers}, we present five particles with the same average jump length (three of them exhibit power law behavior, and two of them follow a Gaussian velocity distribution). All particles start from the same point and are left to explore the space for a thousand iterations. The particles performing Lévy flights exhibit jumps with probability density function $\sim x^{-1-\alpha}$ and an average jump of size $\alpha/(\alpha-1)$ for $\alpha > 1$ (see Pareto distribution \cite{Kleiber2003}). The ones performing Gaussian random walk exhibit jumps with a Gaussian probability density function ($\mu = 0$, $\sigma = 1$) and an average jump of size $\sqrt{2/\pi}$. We make sure that the average jump ($\bar{J}$) is the same for every particle by multiplying the individual jumps with a normalization constant equal to $\bar{J} * (\alpha-1)/\alpha$ for the power laws, and equal to $\bar{J} * \sqrt{\pi/2}$ for the Gaussian jumps. The particles that perform Lévy flights explore the space in a completely different way than the Gaussian ones. Reconsidering platelet transport, how fast platelets reach the vessel walls is a very critical information, and inextricably associated to the underlying characteristics of the probability density function.

This paper is organized as follows. In section \ref{sect:methods} we
describe our methodology and the numerical tools we have developed to
detect the non-Gaussian behavior of PLT motion. We also explain how we
can bridge the scales between the high fidelity blood flow
simulations, which are extremely computationally demanding, and the PLTs
displacement in the impact-R device. Our results are then presented in
section \ref{sect:results}. They consist in a careful statistical
analysis of the random motion of the simulated PLTs (coming from the fully resolved cellular blood flow simulations), and a simulation
of the platelet deposition process of the impact-R (stochastic model), based on the inferred properties of the random walk. Finally, in
section \ref{sect:conclusion} we discuss some implications of our
disrupting theory of PLT motion.

\section{Methods}\label{sect:methods}
Our goal is to understand whether the discrepancy found in Chopard et
al. \cite{Chopard2017} between the Zydney-Colton diffusion
coefficient and the one accounting for the PLTs deposition pattern
observed in the impact-R device can be explained by an anomalous
diffusion process, thus ruling out the possibility to use
eq. (\ref{eq:diffusion}). As discussed in the previous section we made
the hypothesis that PLTs may obey a fat-tailed distributed random walk
instead of a Gaussian one. Due to the great difficulty to measure
directly the movement of platelets (individual trajectories) in whole blood with an in-vitro
approach, numerical simulations are considered. We refer to these
simulations as DNS (Direct Numerical Simulations) as they provide a
high fidelity description of blood, integrating RBCs and PLTs as
deformable suspensions in flowing plasma. However, state of the art
DNS blood solvers are still limited to rather small spatio-temporal
scales.  For this reason we will have to perform a very careful
statistical analysis of the platelet trajectories to evidence their
non Gaussian behavior at the reachable scales.

\subsection{High Fidelity Blood Flow Simulations}\label{sect:DNS}
Here we consider the Palabos-npFEM DNS blood
solver \cite{Kotsalos2019, Kotsalos2020}. It offers high
accuracy, flexibility and high scalability on the top fastest parallel
supercomputers. This computational framework couples the lattice
Boltzmann solver Palabos \cite{PalabosArticle}
for the simulation of blood plasma (fluid phase), a finite element
solver for the resolution of the deformable blood cells (solid phase),
and an immersed boundary method for the coupling of the two
phases. The framework resolves blood cells like RBCs and PLTs
individually (both trajectories and deformed state), including their
detailed non-linear viscoelastic behavior and the complex interaction
between them. 

Collisions between blood particles, whether RBCs or PLTs, are
implemented through a repulsive force acting as a spring, when the
surfaces delimiting two particles are getting too close to each
other. In the current study, we employ the same parameters as reported
in Kotsalos et al. \cite{Kotsalos2019, Kotsalos2020}, where one
can find a detailed description of the numerical models. However, the
intensity of the repulsive force is varied here to account for the extra
repulsion between RBCs and activated platelets whose negative charge is
increased with respect to that of non-activated platelets.

We consider simulations in a 3D box as illustrated in the inset of
Fig. \ref{fig:exp_and_geometry}. The $y$-axis is oriented
vertically.  Two horizontal no-slip walls are positioned at locations
$y=0$ and $y=L$, with $L\in\{ 50~\mu m, 100~\mu m, 250~\mu m, 500~\mu m\}$. A shear flow is produced in the $z$-direction by moving the upper wall
at a proper velocity. The shear rates we consider are
$\dot{\gamma}=100s^{-1}$ and $\dot{\gamma}=400~s^{-1}$ (correspond to the order of magnitude for which thrombosis is observed to occur in cerebral aneurysms \cite{Chopard2017,Boudjeltia2020}). Periodic
boundary conditions are imposed along the $x$ and $z$ horizontal
axes. This setup is meant to approximate the impact-R geometry for
which $L=820~\mu m$ and the $x$ and $z$ axes span a window of size
$1~mm\times1~mm$, embedded in a cylinder of diameter $6.5~mm$
(see \cite{Chopard2017} for more details). The hematocrit is
$H=35\%$ and the ratio RBCs/PLTs is around 5 (substantially larger than the physiological one, though a deliberate choice intended to provide more samples for the statistical analysis of platelet transport).

The simulations are run for a time interval $[t_0,t_1]$, which
typically lasts $1~s$ of physical time. This interval is resolved at
the scale $\sim 10^{-8}~s$, the time step of the Palabos-npFEM
solver. The trajectories $y_i(t)$ along the $y$-axis of all PLTs $i$
are recorded, based on the position of their center of mass. From
theses trajectories, one has to determine the probability distribution of PLT
velocities. To properly sample the time series $y_i(t)$ of positions, one has to
extract the characteristic platelet mean free time, $\dt$, representing the average time
between successive impacts/collisions. This characteristic time is
important, because an under-sampling could result in missing important
collision events, and thus misinterpreting the motion of platelets. On
the other hand, a sampling at a too fine time scale will sample
statistically dependent velocities (ballistic regime), before collisions could randomize the platelet movements (diffusive regime).

Multiple researchers define the sampling interval as $\sim\dot{\gamma}^{-1}$, but this formula has no robust explanation.
Alternatively, we consider the trajectories $y_i$ as seen at different
time scale $\dt$, for several possible values of $\dt \in \{0.01~ms,
0.1~ms, 1.0~ms, 10.0~ms\}$. We  compute the area $A$ formed
graphically by the trajectory $y_i(t)$ at full DNS resolution, namely
\begin{equation}
  A = \int_{t_0}^{t_1} y_i(t) dt
\end{equation}
Then we compute the area formed by the trajectory at scale $\dt$
(i.e. $y_i(t)$ with $t$ incremented by steps $\dt$). These two areas
should be identical if the platelet did not experience a collision
during time $\dt$. The larger $\dt$ for which this equality holds is chosen as the mean free time.

\begin{figure}[H]
    \centering
    
    \begin{subfigure}{0.85\textwidth}
    \centering
    \includegraphics[width=\textwidth]{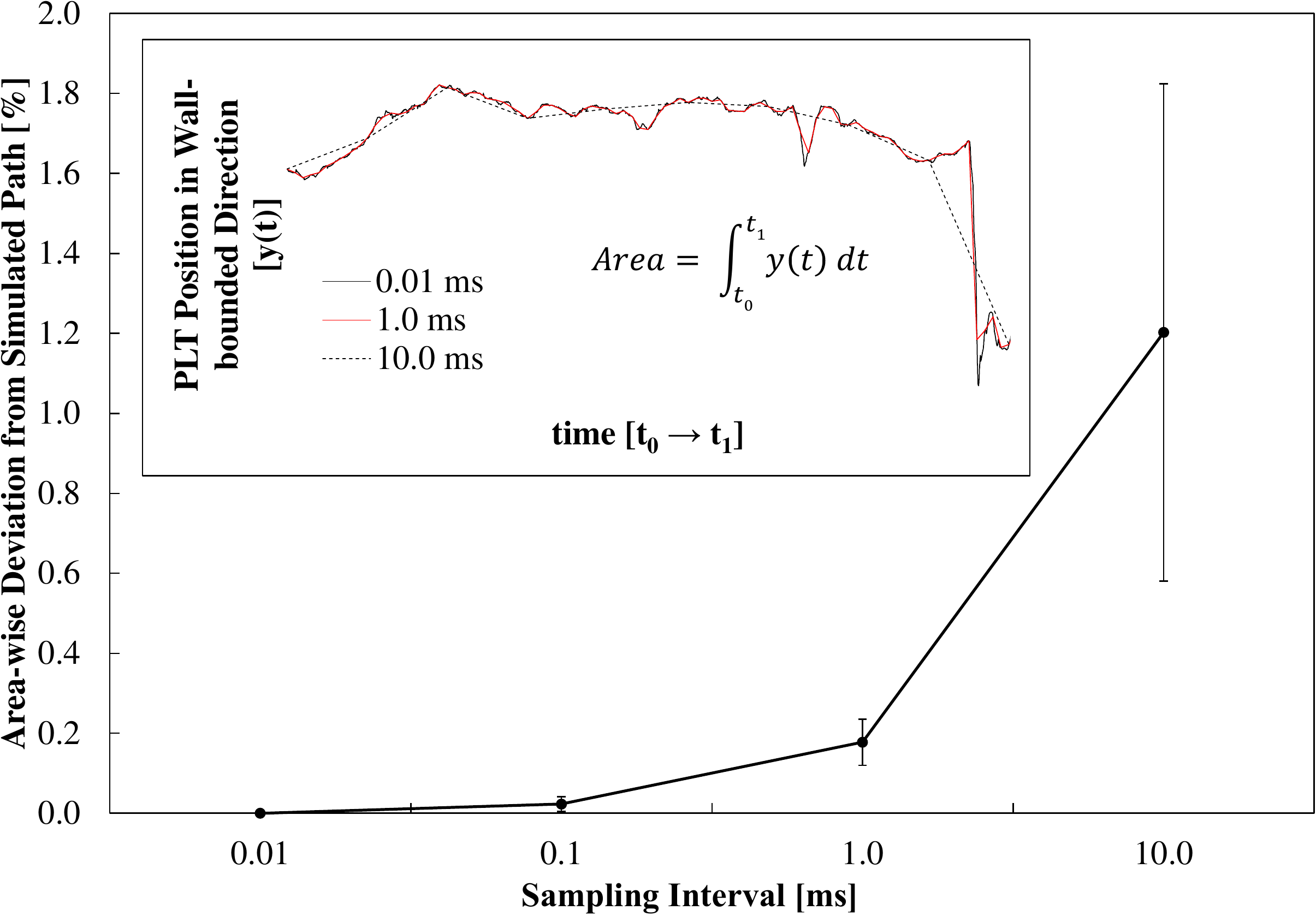}
    \caption{}
    \label{fig:MFP}
    \end{subfigure}
    
    \begin{subfigure}{0.85\textwidth}
    \centering
    \includegraphics[width=\textwidth]{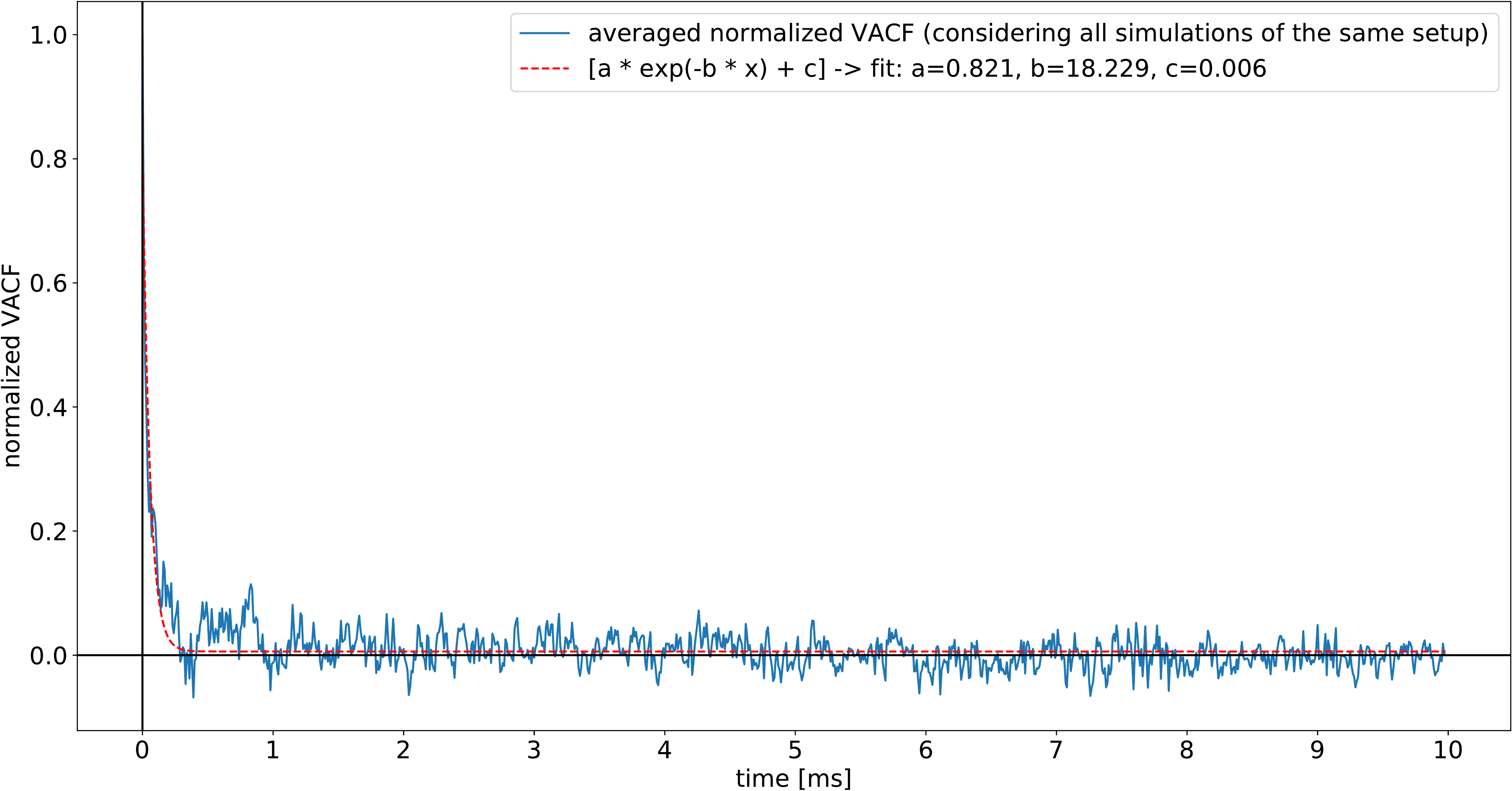}
    \caption{}
    \label{fig:VACF}
    \end{subfigure}
    
    \caption{\textbf{Estimation of the sampling time.} \textbf{(a)} Deviation (area-wise) from the DNS platelet trajectories as a
      function of the time scale at which these trajectories are
      observed. The way the deviation is measured is illustrated in
      the inset, for three different sampling intervals. The black
      circles in the main figure show the average deviation, computed
      over all platelet trajectories and simulations (DNS). The ``error'' bars mark the minimum and maximum deviations.  One observes that a sampling
      interval $\dt=10.0~ms$ (dashed line in the inset) misses many
      collision events, but that $\dt=1.0~ms$ is a proper choice.
      \textbf{(b)} Normalized Velocity Autocorrelation Function averaged over all the simulations of the same setup (scaled impact-R device). The regime after $1~ms$ presents uncorrelated velocities (diffusive regime), while below this threshold there is the ballistic regime (non-zero VACF). Our sampling at $\dt=1.0~ms$ provides uncorrelated velocity samples.}
\end{figure}

Fig. \ref{fig:MFP} presents this method. The vertical axis shows the
relative deviation between these two areas, averaged over all
PLTs and simulations. The inset shows the trajectory of one representative platelet at three different time scales. We see that $\dt=1~ms$ captures well the
change in the trajectory due to collisions, while filtering out the ballistic regime. {\color{black} The same outcome can be retrieved from the velocity autocorrelation function (VACF) defined as
\begin{equation}
  \VACF(t)= \left\langle V_i(t)  V_i(0) \right\rangle
\end{equation}
where $V_i$ is the DNS velocity of PLT $i$ at the wall-bounded direction, and the average is over all platelets of the bulk. As shown in Fig. \ref{fig:VACF}, non-zero VACF (correlated velocities) mark the ballistic time scale, while zero VACF denotes the diffusive regime (random walk). Both methods (Fig. \ref{fig:MFP} \& \ref{fig:VACF}) present matching results, confirming that our sampling does not result in correlated velocities.}

The set of independent platelet velocities is then obtained from the
quantities (where we set $t_0=0$ for simplicity)
\begin{equation}
  v_i(k)={y_i((k+1)\dt)-y_i(k\dt) \over \dt}
\end{equation}
for all platelets $i$.

A classical property characterizing the movement of platelets is
their mean square displacement (MSD), defined as
\begin{equation}
  \MSD(t)=\left\langle ( y_i(t) - y_i(0))^2 \right\rangle
\end{equation}
where the average is over all platelets $i$ of the bulk. This
relation can also be written as
\begin{equation}
    \MSD(t)=\left\langle \sum_{k=1}^{t/\dt}v_i(k)\sum_{\ell=1}^{t/\dt}v_i(\ell)\right\rangle
\end{equation}
Assuming that $v_i(k)$ and $v_i(\ell)$ are independent when $k\ne
\ell$, and that $\langle v_i\rangle=0$ due to the symmetry along the
$y$ axis (fact confirmed numerically for the current simulation setup), we obtain
\begin{equation}
    \MSD(t)=\sum_{k=1}^{t/\dt}\left\langle v_i^2(k)\right\rangle
           ={t\over\dt}\left\langle v_i^2\right\rangle
\label{eq:MSD}
\end{equation}
The second equality comes from the fact that in a steady state, the
velocity distribution is expected to be independent of time. Note
however that this hypothesis is only valid in the bulk of the sample,
where RBCs are found to have a constant density along the $y$-axis. This is no longer the case in the so-called cell-free layer, near the walls at $y=0$ and
$y=L$. For this reason, we separate the treatment of platelets according to their spatial location and determine the velocity distribution for the PLTs in the homogeneous region, away from the walls.

If PLTs encounter no other blood cell, i.e traveling ballistically, then the distance they travel would be proportional to the time interval (distance equals velocity times time), and the MSD would increase quadratically with time. In denser suspensions, quadratic behavior holds only for a very short time interval, of the order of the mean collision time. Beyond this time the motion is better described as a random walk, for which the MSD increases only linearly with time. The rate of growth of the mean square displacement depends on how often the cells suffer collisions. Equation (\ref{eq:MSD}) shows that $\MSD(t)$ is expected to be
proportional to $t$, but this is only the case if $\langle v^2_i \rangle$ (variance of PLT velocities) is finite
and well defined. We will show below that this is not the case for
platelets in a shear flow. Actually the subtle part of this
observation is that, for any numerical simulation, $\langle v^2_i \rangle$ is
obviously finite as there is a finite number of platelets in the
system, for a finite number of time steps. But as the system size is
increased, we observed that $\langle v_i^2 \rangle$ also increases,
indicating a divergence of the variance of platelet
velocities. The striking result is however that for small systems,
namely those that are accessible to DNS blood flow
simulations \cite{Crowl2011,Vahidkhah2014,Mehrabadi2015}, the
non-converged computed value of $\langle v_i^2\rangle$ is compatible
with the Zydney-Colton diffusion coefficient, thus supporting a
misinterpretation of a Gaussian random walk.

\subsection{Statistical Analysis of Platelet Velocities}
In this section we explain how we determined the properties of the
probability distribution of PLT velocities sampled from the high
fidelity blood flow simulations described in section \ref{sect:DNS}.

To avoid wall effects, we discard the velocities corresponding to PLTs
that are less than one RBC diameter from the walls. In order to prove
that PLT velocities are not normally distributed but have a rather
high probability for extreme events, we devised a set
of methods originating from analyses of fat-tailed distributions for
discerning and quantifying power law behavior in empirical data
(in-depth analysis by Clauset et al. \cite{Clauset2009}). A
graphical explanation is provided in the Supplementary Material: Statistical Analysis.

\subsubsection*{Family of Distributions}
The term ``Thick Tails'' is often used to describe distributions with
much higher kurtosis than the Gaussian one, and ``Fat Tails'' is
reserved for both extreme thick tails or membership in the power law
class. We avoid designations such as ``Heavy Tails'' or ``Long Tails''
to keep ambiguity to a minimum \cite{Taleb2020}. The term sub-exponential
distributions is used for the ones that decay more slowly than an
exponential, i.e. they are not exponentially bounded (popular
distribution belonging to this class is the log-normal). To summarize,
the degree of thick tailedness (ranking by severity) is \emph{Thick Tailed} $\supset$ \emph{Sub-exponential} $\supset$ \emph{Power Law/Fat-Tailed (Paretian)}. Furthermore, the thick tailed and
sub-exponential families have all their moments finite. We are
particularly interested in the power law class to describe 
PLT transport, because these distributions  do
not have all their moments finite/defined. This happens for power law
distributions whose tail decays like $\sim x^{-\alpha-1}$ with $\alpha
\leq 2$ (infinite/undefined variance and moments above). For $\alpha
\leq 1$ the mean of the distribution (first moment) is infinite/undefined, as well. 

In statistics, we can never prove that a given distribution describes
the investigated data, instead we can increase our confidence for a
family of distributions against others. In our case, since we
hypothesize that platelet velocities follow power law distributions
with exponent $\alpha \leq 2$, we focus on finding evidences
supporting this null hypothesis. Commonly used distributions that
exhibit power law behavior (asymptotically) are the Pareto, Cauchy
(half-Cauchy, for positive only data), Lévy, Dagum (or Burr Type III),
Singh–Maddala (or Burr Type XII), Log-logistic (or Fisk), inverted
Weibull (or Fréchet) (see Supplementary Material: Power Law Distributions)
\cite{Kleiber2003}. Our approach is to check, for each DNS, the
plausibility of these distributions, according to the fitting criteria
presented below. Also we want to extract the asymptotic power law
exponent and check how many of these distributions exhibit $\alpha \leq 2$.
{\color{black} A question that often arises is why the Pareto alone, a pure power law, is not enough to characterize the data. Indeed, the need to consider more distributions is because Pareto fits the data fairly well toward the higher velocities, but the fit is poor toward the lower values. Thus, considering other distributions with more parameters leads to a better and unbiased fit. Afterwards, the asymptotic analysis of every distribution results in a more accurate estimation of the power exponent $\alpha$ (see the Supplementary Material: Power Law Distributions).}

The distributions whose plausibility lowers our confidence on power
laws are the normal and exponential distributions. Generally, if a
power law is not a better fit than an exponential distribution, there
is scarce ground for considering the distribution to be thick-tailed
at all, let alone a power law \cite{Alstott2014}. Regarding the
sub-exponential class, a careful investigation of the sample moments
can help us decide whether to disregard it or not.

\subsubsection*{Distribution Fitting}
In practice, few phenomena obey power laws for all values of $x$
(where $x$ denotes PLT velocities). More often the power laws apply
only for values greater than some minimum value, $x_{min}$. In such
cases, we say that the tail of the distribution follows a power law
\cite{Clauset2009}. Thereby, for every distribution mentioned
above (power law or not), we find this lower bound and perform the
fitting at the tail.  Regarding the remaining part of the data (body),
this can be sufficiently described by the empirical distribution
function (histogram - see Supplementary Material: Statistical Analysis). Note however that this procedure does not apply when
fitting the data to a Gaussian. In this case, the fit is done across
the whole range of velocities.

Once this lower bound $x_{min}$ is known, every distribution is fitted on the
tail using the Maximum Likelihood Estimate (MLE) method, and its
plausibility is checked using the Kolmogorov-Smirnov (KS) test for
goodness-of-fit. Given the symmetry of the studied phenomenon (for the current simulation setup), we
use the absolute value of the velocities (no discrimination between upward
and downward motions of PLTs). Thus all the statistics are
one-tailed. If the p-value of the KS test is greater than the
significance level (10\% throughout the study for goodness-of-fit
tests), then the investigated distribution could be a plausible fit to
the data.

In addition to this goodness-of-fit test, we employ the Log-likelihood
ratio (LLR) test \cite{Clauset2009} comparing the fitted
distribution with every alternative distribution proposed above (see
Family of Distributions), on the same data. In that case, the smaller the p-value of the
LLR test, the more confident we can be on which distribution is a good
fit of the data. The significance level for the LLR test is set to
1\%. Note that, to confirm or deny a distribution, a small p-value is
``good'' for the LLR test (shows how trustworthy is the test's
result), while it is ``bad'' for the KS test (shows that the
distribution is a poor fit to the data).  Thus, for the
acceptance/plausibility of a distribution fitted on the tail, there
exist two criteria to meet, i.e. the KS and LLR tests.

The estimation of $x_{min}$ is an optimization process based
on minimizing the distance (d) between the investigated model and the
empirical data \cite{Clauset2009}. The metric $d$, that
quantifies this distance, is the widely used Kolmogorov–Smirnov
statistic, which is simply the maximum distance between the Cumulative
Distribution Function (CDF) of the sampled DNS data (the empirical
CDF, noted $S$), and the CDF $P$ of the fitted model. Thus
\begin{equation}
  d = \underset{x\geq x_{min}}{max} \left | S(x) - P(x) \right |
\end{equation}
Therefore, the selected $x_{min}$ is the one that minimizes this
distance. Keep in mind that for every $x$ tested, the parameters of $P$
are fitted using the MLE method, which requires large enough
samples. Thus, the minimization process described above stops when the
remaining tail has less than 100 velocities.

\subsection{Bridging the scale: a Random Walk description}
As indicated previously, we want to show that the unexpectedly high
transport of PLTs observed by Chopard et al. \cite{Chopard2017} in
the in-vitro impact-R device is compatible with the velocity probability
distribution extracted from our DNS blood simulations. A direct
verification is not possible as the spatio-temporal scales
corresponding to the impact-R are still too hard to reach, even on
the fastest supercomputers: a cubic millimeter of blood simulated for
20 seconds.

A solution to bridge this gap of scales is to disregard the detailed
movement of RBCs at the level of the impact-R, and only consider the dynamics
of PLTs in terms of a random walk, using the velocity distribution
obtained from the statistical analysis of the fully resolved PLT
trajectories in the DNS simulations. Therefore we will simulate the
PLTs deposition process taking place in the impact-R through a
stochastic model implementing the determined random motion of
PLTs. The question is to check whether this mesoscopic transport
process reproduces the number of PLTs observed to deposit at the
bottom surface of the impact-R, after $20~s$ of operation. More
precisely, the experimental data \cite{Chopard2017} show that about 3000 activated
platelets per micro-liter of whole blood have disappeared from the bulk
within these first $20~s$.

This method follows the approach of Chopard et
al. \cite{Chopard2017}, by replacing the 1D diffusion equations
describing the bulk of the impact-R device with actual random walks of
point particles. The PLTs that cross the lower boundary are considered
as deposited.  Therefore, for every DNS, we perform the statistical
analysis, and for every candidate of the PLTs velocity distribution, we
simulate the corresponding stochastic model and record the number of
deposited PLTs it predicts after $20~s$.  For more details on
the random walks, see Supplementary Material: Statistical Analysis (Graphical Explanation) \& Stochastic Model Description.

\section{Results \& Discussion}\label{sect:results}
Our analysis builds upon the research by Chopard et
al. \cite{Chopard2017}. In more details, in the in-vitro part of
their study the researchers used the impact-R platelet function
analyzer to study the evolution of PLT deposition
(adhesion-aggregation processes) on the substrate of the
device. Impact-R is a cylindrical apparatus (L=820 $\mu m$ height),
whose lower end is a fixed disc (deposition surface, 132.7 $mm^2$),
and its upper wall is a rotating disc. Due to the rotating upper wall,
the blood is subject to a pure shear flow. The imposed constant shear
rate was 100 $s^{-1}$ (inside an observation window of $1 \times 1$
$mm^2$), and the blood was extracted from seven healthy donors.  The
differential role of activated and non-activated platelets was
analyzed, as well as the role of albumin in the deposition
process. The in-silico counterpart of their study consisted of 1D
diffusion equations describing the movement of activated platelets
(AP) and non-activated platelets (NAP) in the bulk of the device, and
of stochastic rules for PLT deposition on the substrate. The study
revealed that the Zydney-Colton \cite{ZydneyColton} shear induced
diffusion coefficient ($D$) was significantly too small to explain the
observed deposition rate.

The fully resolved 3D cellular blood flow simulations (DNS, section \ref{sect:DNS}) provide a
great amount of information, that simplified mechanistic models or
in-vivo/vitro experiments cannot provide. Numerically following
individual particles and resolving the complex interactions, helped us
develop an alternative theory on how platelets are transported. We
tried to reproduce numerically the in-vitro experiments performed in
Chopard et al. \cite{Chopard2017}, to the extent possible, given
the high computational cost. To reduce the computational demand, we
designed our numerical simulations in channels with lateral dimensions
of 50 $\mu m$, while resolving the wall-bounded direction at L=\{50,
100, 250, 500\} $\mu m$. Consider that the in-vitro counterpart
consists of 1000 $\mu m$ in the lateral directions, and 820 $\mu m$ in
the wall-bounded direction. Numerically, we generated a constant shear
rate flow regime at 100 $s^{-1}$, realized with the help of a moving
top wall and a fixed bottom wall. Periodic boundaries were applied in
the flow and vorticity directions, and the hematocrit was 35\%, as in
the experiments.

Regarding the platelet size and shape, we considered numerical 
experiments with either activated or non-activated PLTs. The NAP are
simulated as nearly-rigid oblate ellipsoids with diameters \{2.5,
3.6\} $\mu m$, thicknesses \{0.6, 1.1\} $\mu m$, and volumes \{2.0,
7.0\} $fl$, respectively. The AP are simulated as nearly-rigid spheres
with diameters \{3.0, 3.6, 4.0, 5.0\} $\mu m$ (covering the
uncertainty from the complicated shape transformations), and volumes
\{12, 22, 30, 60\} $fl$, respectively. Upon PLT activation, negatively
charged phospholipids are translocated from the inner membrane to the
external surface, leading to a more negatively charged PLT. This
complex electrochemical behavior is quantified by the electrophoretic
mobility of platelets \cite{Jy1995,Betts1968,Hampton1966}. Additionally to the change
of the electrophoretic mobility, there is a severe change in shape
with the appearance of blebs and pseudopods, which increases the
hydrodynamic volume of AP. While we address the latter alteration
through the spherical shape, the complex electrochemical behavior is
roughly resolved through an increase in the intensity of the collision
potential between activated PLTs and RBCs, i.e. increased repulsive
forces ranging from 5 to 10 times compared to the ones of NAP.

As for the RBCs, the normal biconcave shape is used in the majority of
the experiments, while in few of them we introduced spherised RBCs
emulating pathological conditions \cite{Boudjeltia2020},
e.g. diabetes, chronic obstructive pulmonary disease.

We performed 64 simulations on Piz Daint, the flagship system of the
Swiss National Supercomputing Center, which is one of the fastest
supercomputers in the world. These
simulations include for completeness, 5 case studies in a tube of
50 $\mu m$ diameter, 5 case studies at 20\% hematocrit, and 6
case studies with constant shear rate at 400 $s^{-1}$. Forty-one
simulations follow the ``exact'' same setup as the experiments of
Chopard et al. \cite{Chopard2017}, but all 64 present
qualitatively the same platelet transport behavior. Out of the 41
simulations, 22 deal with AP (12 of them include repulsive forces),
and the rest deal with NAP. The graphs and results presented below
include these 41 simulations.

\subsection{Anomalous Transport Manifested in Geometry}
\begin{figure}[h]
    \centering
    \includegraphics[scale=0.60]{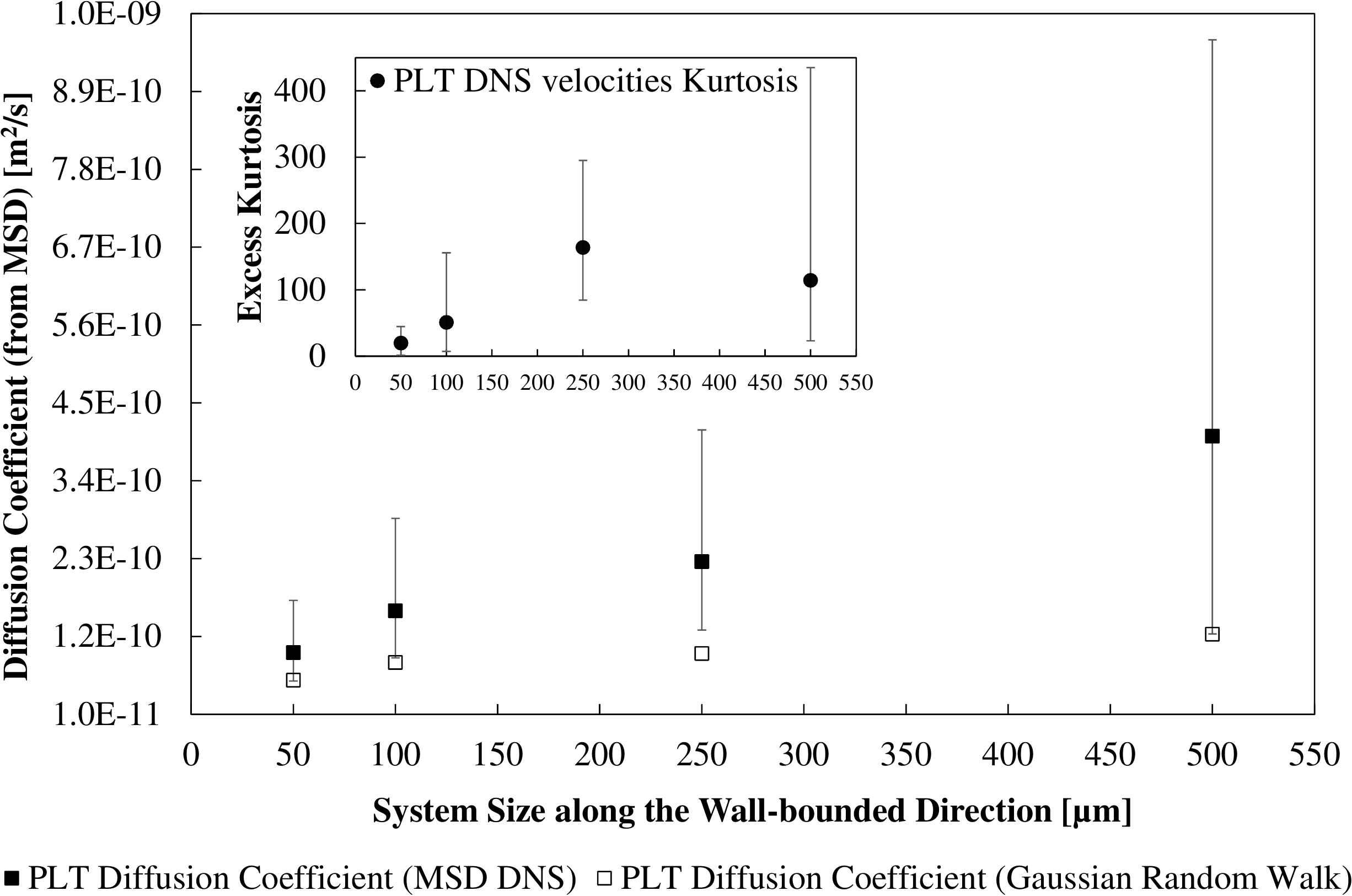}
    \caption{\textbf{Diverging moments indicate fat-tailed distributions.} PLT diffusion coefficient $D$ (black squares) estimated
      from the mean square displacement of PLTs, computed from the
      DNS. The values of $D$ are given as a function of the size $L$
      along the $y$ axis obtained by an average over all the
      simulations. The ``error'' bars denote the minimum/maximum
      coefficients among the various DNS. The white square indicates
      the diffusion coefficient that would correspond to a Gaussian
      random walk by forcing a normal distribution fit on the measured
      PLT velocities. The inset presents the mean excess kurtosis of PLT
      velocities coming from the DNS, as a function of $L$. The
      ``error'' bars correspond to the min/max values of the
      kurtosis. The diverging moments clearly indicate fat-tailed
      distributions with infinite/undefined moments (power law
      behavior with exponent $\alpha \leq 2$).}
    \label{fig:divergingMomements}
\end{figure}

A standard diffusive process is unaffected by the size of the system
(at least away from the walls), i.e. the moments (e.g. variance, kurtosis) of PLT
velocities should converge as the sample size
increases. Fig. \ref{fig:divergingMomements} presents the diffusion
coefficient $D$ of PLTs (AP/NAP) when extracted from the mean squared
displacement (sampled DNS), for channels of varying sizes. Traditionally, the
diffusion coefficient is linked to the MSD, as $D_{MSD}=\MSD(t)/(2t)$ for 1D
systems, e.g. along the wall-bounded direction. The black square
points in Fig. \ref{fig:divergingMomements} indicate the value of
$D_{MSD}$ as a function of the system size, averaged over all DNS simulations.
The ``error'' bars denote the minimum/maximum coefficients among the
various DNS.

In Fig. \ref{fig:divergingMomements} we also evaluate the PLT
diffusivity $D_{Gauss}$ (empty squares), when a Gaussian
distribution is (poorly) fitted on the observed PLT velocities. The
$D_{Gauss}$ is computed from the standard deviation of the corresponding normal
distribution. As expected, the latter approach returns a diffusion
coefficient that is not affected by the system size (small variations
are due to different parameters in the DNS). However, for the
diffusivity emerging from the MSD, there is a diverging behavior as
the sample sizes increase, a signature of a fat-tailed velocity
distribution (diverging second central moment/variance with which MSD is linked to, section \ref{sect:DNS}).  On top of this observation, we show in the inset of
Fig. \ref{fig:divergingMomements} the mean excess kurtosis (fourth
standardized moment) of platelet velocities, which presents a
diverging behavior as well, and values that are way higher than the
corresponding null value of a normal distribution.

Alongside with observing
the moments of the sampled data, we performed a number of normality tests to
check if a Gaussian distribution is a plausible model for the data. We
deployed the Shapiro-Wilk, D'Agostino's $K^2$ and Anderson-Darling
normality tests to check this hypothesis. As expected, every normality
test consistently rejected this hypothesis.

Summarizing, the extremely high kurtosis ($\gg 0$), the failed
normality tests, and the diverging moments of PLT velocities are the
first signs of fat-tailed distributions, and more specifically they
indicate presence of power laws with exponents less than 2,
i.e. infinite/undefined variance (second central moment) and moments above. A striking
observation is that the resulting diffusion coefficients from our
numerical experiments are not far from the Zydney-Colton model
\cite{ZydneyColton} or other numerical studies
\cite{Vahidkhah2014,Crowl2011,Mehrabadi2015}, i.e. they are
consistently two to three orders of magnitude higher than the Brownian
diffusivity $\mathcal{O}(10^{-13})~m^2/s$. As well accepted, our
simulations confirm the role of PLTs-RBCs collisions to enhance PLTs
transport, but a further analysis implies that the
diffusive process is anomalous. Possibly, the suggestion for
extra drift terms or rheological potentials
\cite{Eckstein1991,Crowl2011,Kumar2012} comes from a
misinterpretation of this anomalous diffusive process.

\subsection{Power Law Emergence}
\begin{figure}[h]
    \centering
    \includegraphics[scale=0.4]{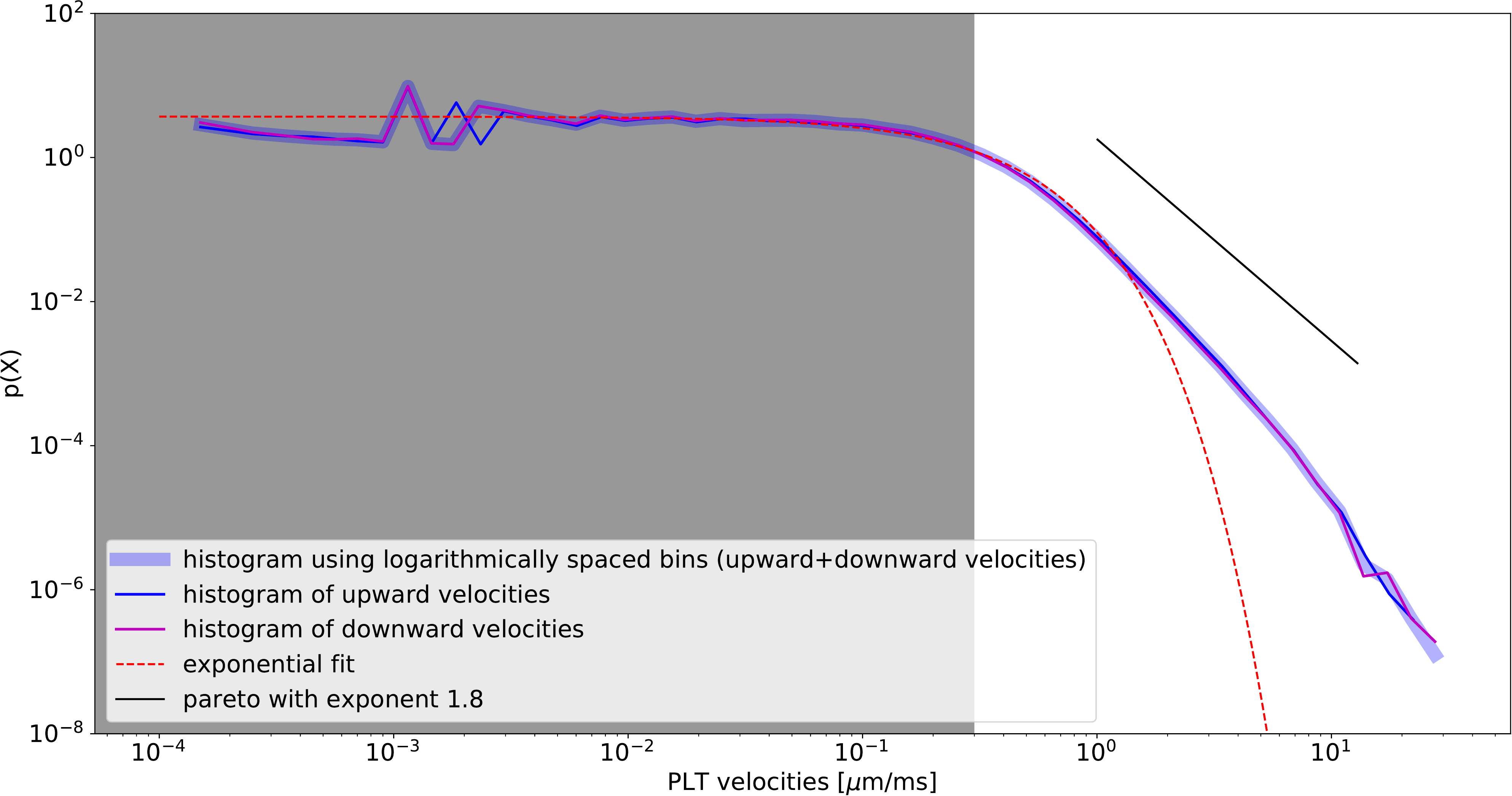}
    \caption{\textbf{Histogram of PLT velocities on doubly logarithmic axes using logarithmically spaced bins.} The shaded area is the body of the distribution, which is of no interest for the current analysis, while the non-shaded area roughly presents the tail of the data. The histogram accumulates the velocities from all the numerical experiments performed, discriminating their upward or downward direction, and thus revealing the symmetry of PLT motion in the current setup (impact-R device). An exponential distribution fails to describe the decay of the tail, while it is evident that a power law fits better on the data.}
    \label{fig:histogram}
\end{figure}

From the statistical analysis of the sampled DNS output, we tried to
find evidences that PLT velocities follow fat-tailed distributions,
and more specifically power laws with exponent $\alpha \leq
2$. Out of the 41 numerical experiments only 2 of them did not
show evidence of our hypothesis, i.e. no valid fitting of power laws on
the data was obtained. For the rest, the tails of PLT velocities can be described with distributions which asymptotically behave as power laws (see
Family of Distributions in section \ref{sect:methods}, and the Supplementary
Material: Power Law Distributions). {\color{black} Fig. \ref{fig:histogram} gathers the PLT velocities from the 41 numerical simulations. The histogram uses logarithmically spaced bins to accurately capture the tail of the distribution. The velocities are discriminated as upward/downward validating the symmetry of the setup, i.e. a PLT has equal chance to move up or down. Given this observation, the statistical analysis is performed on the absolute value of the velocities. Additionally, Fig. \ref{fig:histogram} shows an exponential fitting on the data, which clearly fails to capture the decay of the tail. Indeed, the tail regime (white background of Fig. \ref{fig:histogram}) presents a power law behavior, i.e. straight line in a log-log plot of the histogram. The Pareto curve plotted in Fig. \ref{fig:histogram} is just for visualization purposes, since the statistical analysis should be performed per DNS. As mentioned by Clauset et al. \cite{Clauset2009}, the regression methods on doubly logarithmic histograms give significantly biased exponents, and thus our statistical analysis deploys robust tools, other than plotting \& fitting, to reveal the unbiased power law exponent $\alpha$.}

Table \ref{tab:exp_depo_PLT_type} shows, for the different platelet
types (non-activated platelets NAP, activated platelets AP-no rep, activated
platelets with repulsive forces AP-rep), the power law
exponent $\alpha$ averaged over all the DNS experiments and the fitted/accepted power laws
with exponent $\leq 2$. The min/max indicated in
table \ref{tab:exp_depo_PLT_type} denote the minimum and maximum
values of the exponent from the fitting of the various distributions
(Pareto, half-Cauchy, Lévy, Dagum, Singh–Maddala, Log-logistic, and
inverted Weibull) for each PLT type. A remark is that we are not biased towards the exponents $\leq 2$ (discarding the values above), since our previous observations, i.e. on diverging moments, high kurtosis and failed normality tests, support this direction, and in parallel reject the alternative hypothesis on thin-tailed distributions.

It is interesting to observe that the shape and the electrophoretic
properties of platelets are reflected on the exponent of the power
law. The lower the power law exponent, the higher the mobility through
``extreme'' tail events. The exponents observed for AP-rep are
consistently smaller. In the Supplementary Material: Summary of DNS and Statistical Analysis, we provide a
table that summarizes the majority of the performed experiments with
key quantities per DNS for the completeness of the study, e.g. one can
find per DNS the power law exponents (considering all fitted/accepted
distributions with exponent $\alpha \leq 2$).

\begin{table}[h]
\centering
\caption{\textbf{Power law exponent and deposition rates.} Mean power law exponent for the different PLT types, averaged
  over the simulations and power law distributions (the ones that give $\alpha \leq 2$). The min/max denote
  the variation of the exponent across the simulations. The simulated (random walk model)
  mean number of deposited PLTs in the impact-R is computed
  considering all the simulations and power law distributions (the ones that give $\alpha \leq 2$) per PLT type.  The maximum deposition
  corresponds to the lowest power law exponent (fatter tails). The
  mean deposited PLTs based on a (poor) fitting of normal or
  exponential distributions heavily underestimate the deposition
  observed in Chopard et al. \cite{Chopard2017}.}
\label{tab:exp_depo_PLT_type}
\begin{tabular}{|c|c|c|c|}
\hline
& \textbf{NAP}     & \textbf{AP - no rep} & \textbf{AP - rep} \\ \hline
{Mean Power Law $\alpha$ (min/max)} & 1.56 (1.35/1.75) & 1.48 (1.23/1.75)     & 1.40 (1.12/1.65)  \\ \hline
{Mean Deposited PLTs (power laws)} & 366  & 321  & 416  \\ \hline
{Max Deposited PLTs (power laws)}  & 1759 & 1512 & 2270 \\ \hline
{Mean Deposited PLTs Normal Distribution}      & 246  & 174  & 287  \\ \hline
{Mean Deposited PLTs Exponential Distribution} & 250  & 175  & 290  \\ \hline
\end{tabular}%
\end{table}

In a second step, we used the fitted distributions as generative
mechanisms for simulating random walks and the transport of PLTs in
the impact-R device. Table \ref{tab:exp_depo_PLT_type} presents the
amount of deposited PLTs (the ones that cross the bottom wall) after
running the stochastic simulations for $20~s$ of physical time, with
$L=820~\mu m$. Chopard et al. \cite{Chopard2017} report that 3125
AP have disappeared from the bulk during these $20~s$, with an initial
concentration of 4808 AP per micro-liter. Given that the phenomenon is symmetric along the $y$-axis (cross-checked with the DNS), and due to
the developed cell free layer close to the walls (trapping the
crossing PLTs), we expect at least 1500 platelets out of 4808
($1500 \sim 3125/2$) to cross the bottom wall of the system.

Table \ref{tab:exp_depo_PLT_type} shows that this expected number of
deposited platelets is compatible with the proposed fat-tailed velocity
distribution, without the need to invoke special drift terms or a
rheological potential. The table also presents the amount of
deposited PLTs when forcing a Gaussian or exponential fit on the
velocity distributions. Clearly, the deposition values are way too
small to describe the deposition rates observed in the in-vitro
experiments. Therefore, the use of standard diffusive models heavily
underestimates the deposition rate, compared to the Lévy flights that
produce a 2 to 10 times higher amount of deposited PLTs.

\begin{figure}[h]
    \centering
    \includegraphics[scale=0.60]{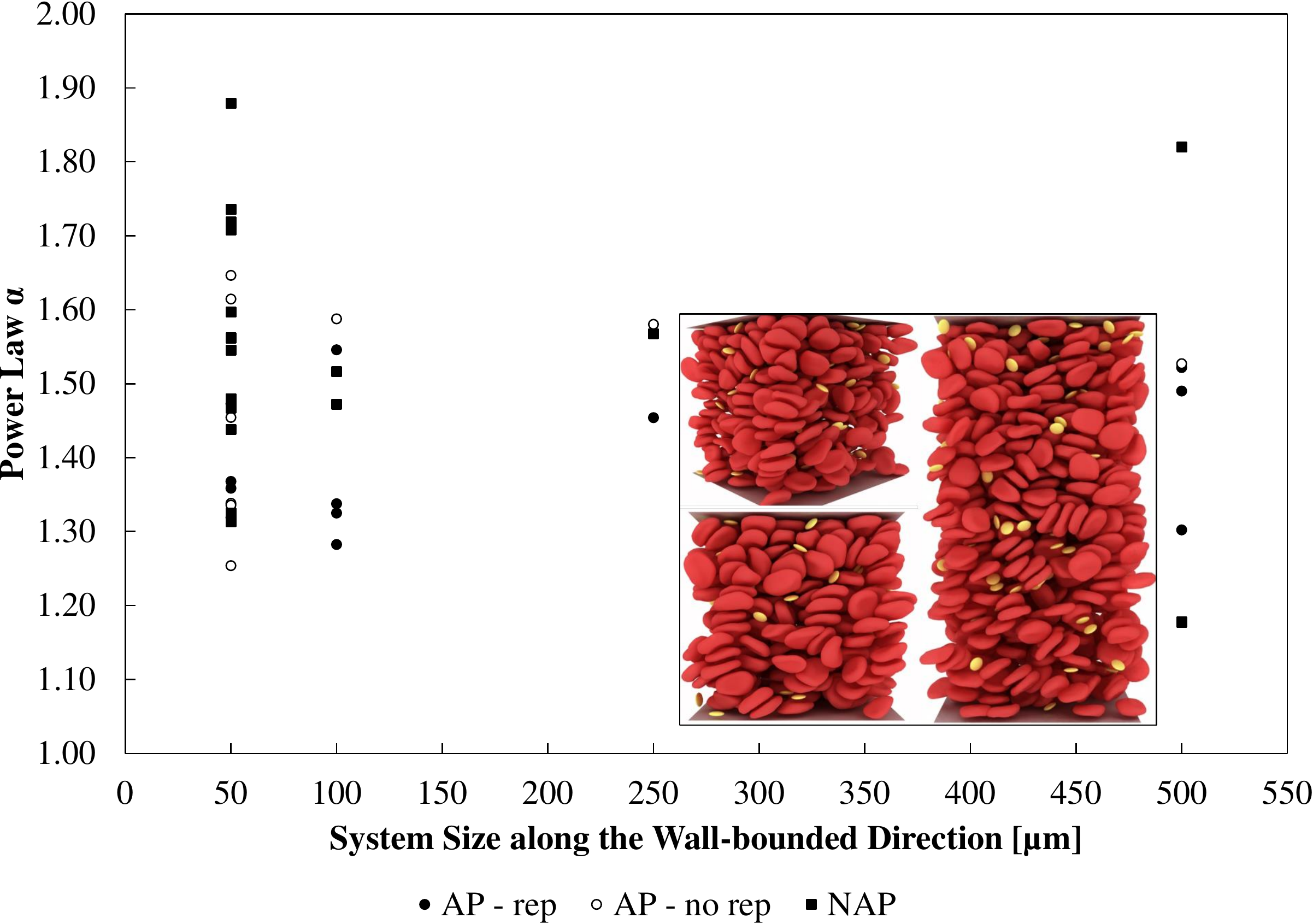}
    \caption{\textbf{Power law exponent invariance.} Mean power law exponent per cellular blood flow
      simulation. The power laws are independent of the system size
      (no diverging behavior as with the moments of the velocity
      distribution). Any variation is due to the different parameters
      per simulation (shapes, sizes, repulsive forces). The inset
      shows a few simulation snapshots from systems of size 50 \& 100
      $\mu m$ along the wall-bounded direction $y$.}
    \label{fig:exp_and_geometry}
\end{figure}

As shown in Fig. \ref{fig:exp_and_geometry}, it is also interesting to
note that as opposed to the MSD which diverges with the system
size, the exponent $\alpha$ keeps a consistent value when varying $L$
from $50~\mu m$ to $500~\mu m$ (scale-invariance of power laws, i.e. the phenomena are expected to occur without a characteristic size or scale).

\section{Conclusions}\label{sect:conclusion}
Combining the power law fitting on the tails with the normality tests
that fail in all the experiments, the diverging moments of the
velocity distribution, the very high kurtosis, and the good agreement
with the impact-R deposition data, we give convergent evidences that PLTs
do not follow a Gaussian random walk, but rather Lévy Flights. Therefore, it is the properties of the system that lead us towards the validation of our hypothesis. The
standard diffusion equation does not apply to describe PLTs
transport. Fractional differential equations might be needed to
account for such an anomalous diffusion process at the macroscopic
scale.

As mentioned by Kumar and Graham \cite{Kumar2012}, no clear and
systematic mechanistic explanation was yet proposed for the
segregation and margination phenomena of platelets. In particular, no
simplified mathematical description (such as a set of transport
equations or a simple stochastic process model) has emerged that
captures the phenomena. In this study, we prove that PLT velocities,
more specifically the tail of their distribution, can be described by
power laws ($P(v) \sim v^{-\alpha-1}$) with exponent $\alpha \leq 2$. We found no evidence
of normally distributed PLT velocities, and thus PLTs cannot possibly
exhibit standard diffusion, which is the norm when describing PLT
transport.

The new stochastic process model that we introduce does not
need additional terms to describe margination, as is the case in other
studies \cite{Eckstein1991,Crowl2011,Mehrabadi2015}. A striking
observation is that while our results  are compatible
with the standard models (diffusion coefficient extracted from MSD), a further investigation (statistical analysis)
reveals the anomalous behavior of PLT transport. This can be
explained from the fact that the more fat-tailed a distribution, the
more statistical information lies in the tail, and the moments (on
which standard diffusive models are built upon) become uninformative
and unreliable. The majority of the numerical research in this field
is limited to case studies that deal with too few blood cells in the
computational domain, and this is due to the high computational cost
of the simulations. Our highly scalable numerical framework allows us
to investigate cases with dimensions approaching the ones of the
in-vitro experiment. We would like to emphasize that the ability to
study sizes of clinical relevance (at least resolving the direction of
interest), allowed us to observe the idiosyncratic behavior of PLT
transport, and to capture the anomalous characteristics that
manifest in setups of larger sizes.

{\color{black} For the current study, based on in-vitro experiments \cite{Chopard2017}, we limited the exploration of the parametric space to $35\%$ hematocrit and low shear rates. Given that platelet transport is a collision-driven phenomenon, we believe that there should be a limit between $0$ (PLT rich plasma) and $35\%$ hematocrit, below which the normal diffusion is restored. We have performed a handful of numerical experiments at $20\%$ hematocrit revealing that anomalous transport is still occurring provided that repulsive forces (activated PLTs) are present. On the other hand, there was no evidence for anomalous transport of non-activated PLTs at this low hematocrit value. More in-vitro \& in-silico experiments are critical to shed light on this new interpretation of PLT transport, and to determine the power law exponent as a function of the shear rate and hematocrit.}

In addition to proposing a disruptive view of PLT transport physics in blood,
our results have a concrete impact on the design of new and efficient
platelet function tests. Those can  be a vital part for the detection of
cardio/cerebrovascular diseases in clinical practice.
Nowadays, there exist numerous platelet function tests for the
diagnosis of disorders or the monitoring of anti-platelet therapies, but with
limited prognostic capacity in clinical practice due to contradicting
results. As several studies show
\cite{Breet2010,Picker2011,Koltai2017}, there is a problem
interpreting the results and mapping them to patient risk and
disease. We strongly believe that the emerging stochastic models from
the present analysis will offer a paradigm shift for developing the
next generation tests, as the understanding of PLT transport is
inextricably associated with the success of platelet function
testing. A first step in this direction was proposed by Dutta et
al. \cite{Dutta2018}, a study in which clinically important
properties, such as platelet adhesion and aggregation rates, can only
be properly inferred provided that PLT transport can be correctly
described.

\clearpage
\section*{Supplementary Material: Statistical Analysis (Graphical Explanation)}
\begin{figure}[h]
    \centering
    \includegraphics[scale=0.40]{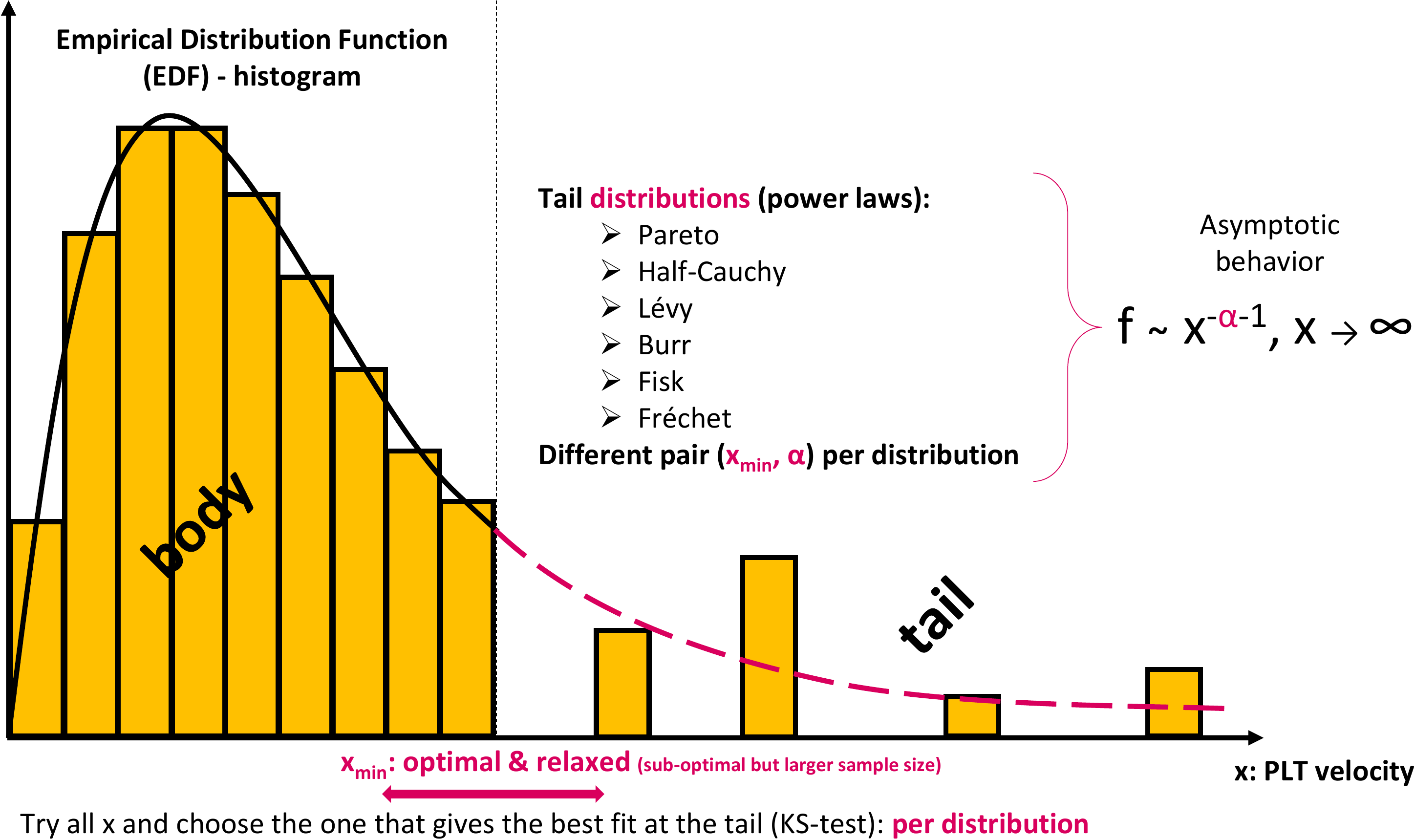}
    \caption{Statistical analysis per direct numerical simulation. Per investigated distribution, we define $\mathbf{x_{min}}$ with the optimization process as described in Methods, and $\boldsymbol{\alpha}$ through the MLE method.}
    \label{fig:statistics_graphical}
\end{figure}

For every cellular blood flow simulation, we produce from the sampled output the histogram of the absolute platelet velocities (velocities along the wall-bounded direction for channel flow, and radial velocities for tubular flow). Our goal is to fit a model on this histogram and use it for simulating random walks. Below some lower bound ($x_{min}$), the original data can be sufficiently described by the empirical distribution function (histogram), but above $x_{min}$, due to the fractionate information that we get from the tail, we need to apply more sophisticated statistical models. Since we are particularly interested on investigating the power law hypothesis, we fit a family of distributions that asymptotically exhibits power law behavior \cite{Kleiber2003}, e.g. Pareto, half-Cauchy, Lévy, Dagum, Singh–Maddala, Log-logistic, and inverted Weibull. Every distribution is fitted separately on the data (just at the tail), and therefore the lower bound and power law exponent $\alpha$ vary per distribution (see figure \ref{fig:statistics_graphical}). The distributions with $\alpha \leq 2$ are characterized as fat-tailed or power laws, and their exponents (presented as mean, minimum \& maximum per DNS) describe the fatness of the tail, i.e. the lower the exponents the higher the tail events (large PLT velocities). This analysis is performed for every cellular blood flow simulation, and the resulting power law exponents are grouped/presented per PLT type (activated, non activated, and with/without repulsive forces).

To generate data for the random walks, the DNS sampled output (per simulation) is split into the body and the tail (see figure \ref{fig:statistics_graphical}). Regarding the body, we use the empirical distribution function (ECDF) for generating new velocities. For the generated velocities that are above the $x_{min}$ (varies per fitted distribution), we use the fitted distributions to re-generate velocities belonging to the tail. For the new data to be as close as possible to the original (from DNS), the fitted distributions should pass the goodness-of-fit and LLR tests. Assuming a plausible distribution for the tail, it may be the case that the tail sample size is small ($n \approx 100$). Fitting on samples of small sizes should be treated with caution, because power laws may appear to be a good fit even when the data are drawn from a non-power law distribution. An additional step to strengthen our confidence on power law behavior is to relax the $x_{min}$ threshold, i.e. reduce the lower bound and thus increase the sample size on which to fit. For this relaxed version, we need to perform an additional goodness-of-fit test \cite{Clauset2009}. For this test, we use as metric the Kuiper's statistic, which is a variation of the Kolmogorov-Smirnov, but more sensitive on capturing differences at the tails. In short, we generate data from the ECDF, then synthetic data with the optimal $x_{min}$ (below $x_{min}$ we use the ECDF, above $x_{min}$ we use the fitted model), and we compare the two data sets using the Kuiper's test, from which we get the statistic. Following, we generate another set of synthetic data with the relaxed $x_{min}$, and we compare it with the data from the ECDF, extracting another statistic. If the last statistic is smaller than the one of the first comparison, then the relaxed model is a valid alternative. By repeating this process more than a thousand times \cite{Clauset2009}, we get a p-value coming from the fraction of valid/accepted relaxed models to the overall repetitions. Keep in mind, that the relaxed tail gives an additional fitting for consideration, thus for each distribution we have the \emph{optimal and relaxed fittings}. Therefore, per DNS and per fitted distribution, we construct a generative mechanism that ``feeds'' the random walk models.

\clearpage
\section*{Supplementary Material: Power Law Distributions}
The probability densities presented here are in the ``standardised'' form, i.e. the location parameter is zero and the scale parameter is one. However, the analysis is not affected by this ``standardised'' form, since these parameters are only for shifting and scaling the investigated distributions, respectively. The power law exponent $\alpha$ is given by the asymptotic analysis of the distributions (probability density function-pdf), which decay like $f \sim x^{-\alpha-1}$. The presentation of the pdf with $-1$ in the exponent is deliberate, since the survival function ($\bar{F} = Pr(X > x)$, complementary cumulative distribution function) decays like $\bar{F} \sim x^{-\alpha}$.

\subsection*{\small Pareto}

The probability density function for Pareto is: $f(x, b) = \frac{b}{x^{b+1}}$ for $x \ge 1$, $b > 0$. For $x \rightarrow \infty$, $f(x, b) \sim x^{-b-1}$. The variance is $\infty$ for $b \leq 2$ and the mean is $\infty$ for $b \leq 1$. The power law exponent is $b$.

\subsection*{\small Half-Cauchy}

The probability density function for half-Cauchy is: $f(x) = \frac{2}{\pi (1 + x^2)}$ for $x \ge 0$. For $x \rightarrow \infty$, $f(x) \sim x^{-2}$. The mean and variance are undefined. The power law exponent is fixed to $1$.

\subsection*{\small Lévy}

The probability density function for Lévy is: $f(x) = \frac{1}{\sqrt{2\pi x^3}} \exp\left(-\frac{1}{2x}\right)$ for $x \geq 0$. For $x \rightarrow \infty$, $f(x) \sim x^{-1.5}$. Both the mean and the variance are $\infty$. The power law exponent is fixed to $0.5$.

\subsection*{\small Dagum (or Burr Type III)}

The probability density function for Dagum is: $f(x, c, d) = c d x^{-c - 1} / (1 + x^{-c})^{d + 1}$ for $x \geq 0$ and $c, d > 0$. For $x \rightarrow \infty$, $f(x, c, d) \sim x^{-c-1}$. The mean is undefined for $c \leq 1$, and the variance is undefined for $c \leq 2$. The power law exponent is $c$.

\subsection*{\small Singh–Maddala (or Burr Type XII)}

The probability density function for Singh–Maddala is: $f(x, c, d) = c d x^{c-1} / (1 + x^c)^{d + 1}$ for $x \geq 0$ and $c, d > 0$. For $x \rightarrow \infty$, $f(x, c, d) \sim x^{c-1} / x^{c d + c} \sim x^{-c d - 1}$. The mean is undefined for $c d \leq 1$, and the variance is undefined for $c d \leq 2$. The power law exponent is $c d$.

\subsection*{\small Log-logistic (or Fisk)}

The probability density function for Log-logistic is: $f(x, c) = c x^{-c-1} (1 + x^{-c})^{-2}$ for $x \geq 0$, $c > 0$. For $x \rightarrow \infty$, $f(x, c) \sim x^{-c-1}$. The mean is undefined for $c \leq 1$, and the variance is undefined for $c \leq 2$. The power law exponent is $c$.

\subsection*{\small Inverted Weibull (or Fréchet)}

The probability density function for inverted Weibull is: $f(x, c) = c x^{-c-1} exp(-x^{-c})$ for $x > 0$, $c > 0$. For $x \rightarrow \infty$, $f(x, c) \sim x^{-c-1}$. The mean is undefined for $c \leq 1$, and the variance is undefined for $c \leq 2$. The power law exponent is $c$.

\clearpage
{\color{black}
\section*{Supplementary Material: Stochastic Model Description}
The Impact-R is a cylindrical device whose bottom wall is a fixed disc (deposition substrate), while the upper wall is a rotating disc (shaped as a cone with a small angle). The height of the device is $0.82~mm$ and due to the motion of the upper wall a pure shear flow is created. A controlled shear rate $\dot{\gamma}$ is produced in a given observation window of $1 \times 1~mm^2$, where we track the formation of clusters resulting from the deposition and aggregation of platelets.

The platelets reach the bottom layer due to a RBC-enhanced shear-induced diffusion. This process is described by a random walk as
\begin{align*}
    \Delta z (t) &= \lambda ~ v_z ~ |s_z| ~ dt \\
    \Delta x (t) &= v_{xy} ~ |s_{xy}| ~ \cos(2 \pi r) ~ dt \\
    \Delta y (t) &= v_{xy} ~ |s_{xy}| ~ \sin(2 \pi r) ~ dt
\end{align*}
where $v_{z,xy}$ are speed units coming from the DNS, $r$ is a random variable uniformly distributed in $[0,1[$, $s_{z,xy} \in ] 0, \infty [$ are random variables distributed according to the fitted models from the cellular blood flow simulations, $\lambda \in \{-1,1\}$ with probability $1/2$ for each outcome (given the proven symmetry of the current setup), and $dt$ is the time step of the simulation. The random motion in the $xy$-plane can be omitted for simplicity. Superimposing the stochastic motion with the velocity field of the pure shear flow, the positions of the platelets are updated as
\begin{align*}
    z_i (t+dt) &= z_i (t) + \Delta z_i (t) \\
    x_i (t+dt) &= x_i (t) + \Delta x_i (t) + \dot{\gamma}~z_i~dt \\
    y_i (t+dt) &= y_i (t) + \Delta y_i (t)
\end{align*}
The AP and NAP crossing the lower boundary of the computational domain are candidates for deposition (trapped in the Cell Free Layer -CFL-, and never re-injected into the bulk), periodic conditions are applied at the $x,y$ directions, and bounce back boundary condition for the platelets that cross the upper boundary. To simplify our analysis and get a rough estimate of the deposited PLTs, we do not consider any deposition rule (as in Chopard et al. \cite{Chopard2017}), but we assume that a PLT deposits when it crosses the lower boundary. Figure \ref{fig:RandomWalk_ProblemDescription} summarizes this stochastic process.

\begin{figure}[h]
    \centering
    \includegraphics[scale=0.45]{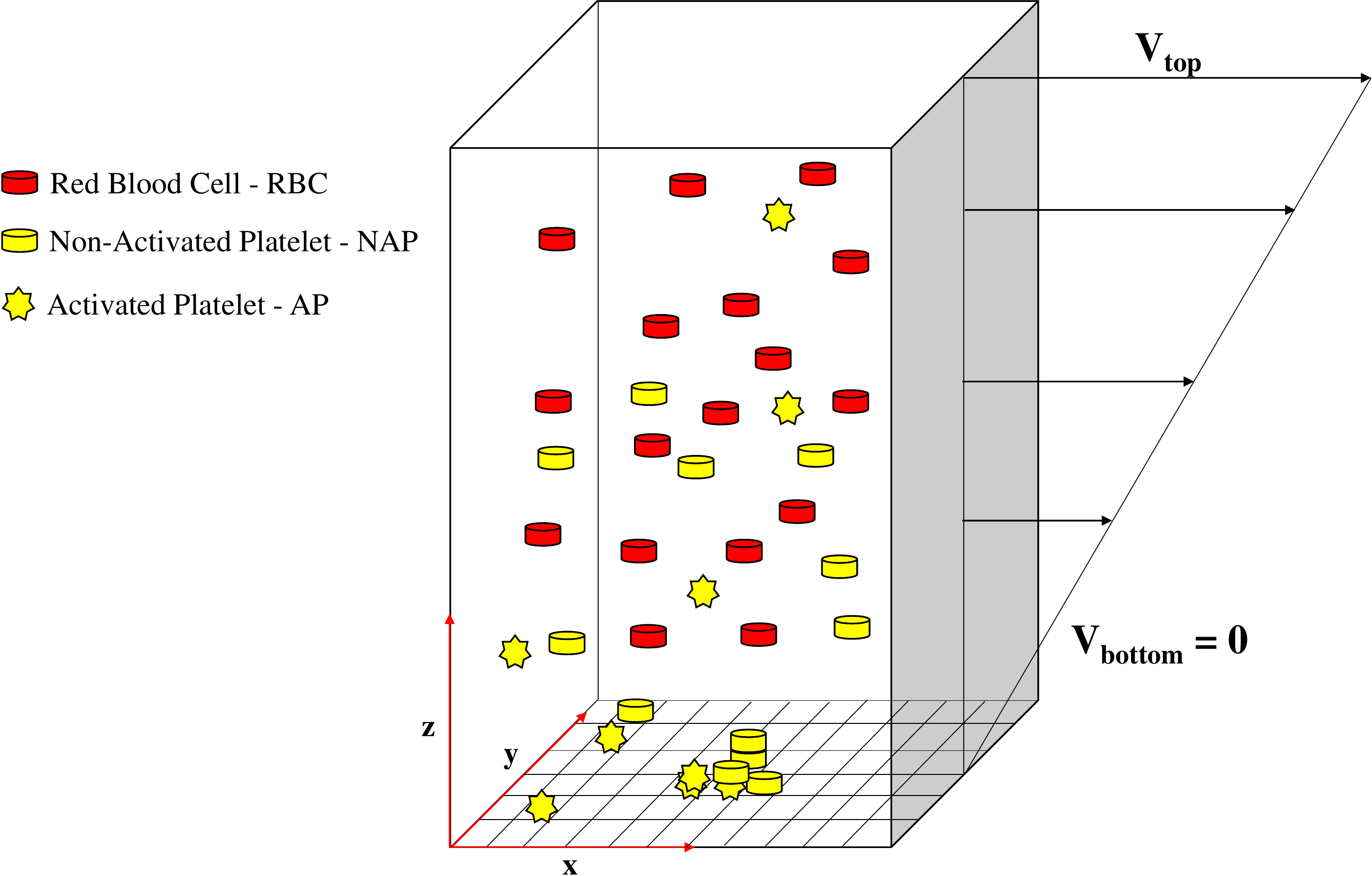}
    \caption{Window of Impact-R device: The bottom wall is a fixed boundary of dimensions $1 \times 1~mm^2$, the wall-bounded direction is $0.82~mm$. The bulk contains whole blood at $35\%$ hematocrit. The discretization of the substrate is such that in every cell can fit just one platelet. The initial densities of the blood particles are determined by the \emph{in vitro} experiment \cite{Chopard2017} and usually are about: $172~200~(\mu l)^{-1}$ for NAP, $4808~(\mu l)^{-1}$ for AP.}
    \label{fig:RandomWalk_ProblemDescription}
\end{figure}
}

\afterpage{
\clearpage
\newgeometry{margin=1.5cm,includefoot}
\begin{table}[]
\centering
\caption*
{ {\Large \textbf{Supplementary Material: Summary of DNS and Statistical Analysis}} \\ \\
{\footnotesize
\emph{PLT}: Activated/Non-Activated platelets, \emph{rep}: no repulsive forces at 20 (weight of collision potential-DNS related), \emph{exp}: either scaled impact-R (box) or tube, \emph{dir}: system size along the wall-bounded direction/diameter [$\mu m$], \emph{Ht}: hematocrit [\%], \emph{SR}: Shear Rate [$s^{-1}$], \emph{num RBCs/PLTs}: number of RBCs/PLTs in DNS, \emph{vel}: mean absolute velocity along \emph{dir} from DNS [$\mu m/ms$], \emph{D (DNS)}: Diffusion Coefficient - MSD of DNS [$m^2/s$], \emph{D (Gaussian)}: Diffusion Coefficient - Random Walk (Gaussian fitted on DNS) [$m^2/s$], \emph{fat-tails}: number of accepted power laws (considering only the ones with $\mathbf{\boldsymbol{\alpha} \leq 2}$) for both optimal and relaxed tail fitting, \emph{min sample size}: minimum sample size of one of the accepted power laws, \emph{avg sample size}: average sample size considering all accepted power laws, \emph{avg exp}: average power law exponent considering all accepted power laws, \emph{max depo}: deposited PLTs (after 20s physical time \cite{Chopard2017}) using the random walks (Impact-R at full scale) corresponding to the \emph{min exp}. Simulations that seem to repeat are with/without the Particle In Kernel (PIK) technique \cite{Kotsalos2020} (validating its consistency).
}
}
\label{tab:exps_summary}
\resizebox{\textwidth}{!}{%
\begin{tabular}{|c|c|c|c|c|c|c|c|c|c|c|c|c|c|c|c|c|c|c|} 
\hline
\textbf{PLT}  & \textbf{rep}  & \textbf{exp}  & \textbf{dir}  & \textbf{Ht}  & \textbf{SR}  & \begin{tabular}[c]{@{}c@{}}\textbf{num}\\\textbf{ RBCs} \end{tabular} & \begin{tabular}[c]{@{}c@{}}\textbf{num}\\\textbf{ PLTs} \end{tabular} & \textbf{vel} & \begin{tabular}[c]{@{}c@{}}\textbf{D }\\\textbf{ (DNS)} \end{tabular} & \begin{tabular}[c]{@{}c@{}}\textbf{D }\\\textbf{ (Gaussian)} \end{tabular} & \begin{tabular}[c]{@{}c@{}}\textbf{excess}\\\textbf{ kurtosis} \end{tabular} & \textbf{fat-tails }  & \begin{tabular}[c]{@{}c@{}}\textbf{min}\\\textbf{ sample}\\\textbf{ size} \end{tabular} & \begin{tabular}[c]{@{}c@{}}\textbf{avg}\\\textbf{ sample}\\\textbf{ size} \end{tabular} & \begin{tabular}[c]{@{}c@{}}\textbf{avg}\\\textbf{ exp} \end{tabular} & \begin{tabular}[c]{@{}c@{}}\textbf{min}\\\textbf{ exp} \end{tabular} & \begin{tabular}[c]{@{}c@{}}\textbf{max}\\\textbf{ exp} \end{tabular} & \begin{tabular}[c]{@{}c@{}}\textbf{max}\\\textbf{ depo} \end{tabular} \\ 
\hline
AP & 200 & box & 500 & 35 & 100 & 4765 & 953 & 0.329 & 3.00E-10 & 1.12E-10 & 23.3 & 2 & 640 & 822 & 1.522 & 1.35 & 1.70 & 367 \\ 
\hline
AP & 200 & box & 500 & 35 & 100 & 4765 & 953 & 0.319 & 3.11E-10 & 1.01E-10 & 31.4 & 3 & 362 & 619 & 1.302 & 1.00 & 1.63 & 384 \\ 
\hline
AP & 200 & box & 50 & 35 & 100 & 476 & 95 & 0.367 & 1.40E-10 & 1.27E-10 & 17.5 & 4 & 181 & 772 & 1.319 & 1.00 & 1.59 & 640 \\ 
\hline
AP & 200 & box & 50 & 20 & 100 & 272 & 54 & 0.218 & 1.54E-10 & 4.87E-11 & 5.4 & 3 & 104 & 540 & 1.314 & 1.00 & 1.49 & 413 \\ 
\hline
AP & 200 & box & 100 & 35 & 100 & 953 & 190 & 0.419 & 2.87E-10 & 1.77E-10 & 26.8 & 2 & 508 & 3081 & 1.325 & 1.30 & 1.35 & 879 \\ 
\hline
AP & 200 & box & 50 & 35 & 100 & 476 & 95 & 0.361 & 1.71E-10 & 1.28E-10 & 39.2 & 8 & 102 & 1194 & 1.359 & 1.00 & 1.62 & 2270 \\ 
\hline
AP & 200 & box & 50 & 20 & 100 & 272 & 54 & 0.237 & 1.11E-10 & 5.65E-11 & 4.6 & 3 & 101 & 880 & 1.735 & 1.50 & 1.86 & 375 \\ 
\hline
AP & 200 & tube & 52 & 35 & 200 & 374 & 74 & 0.337 & 1.06E-10 & 1.07E-10 & 20.0 & 6 & 136 & 402 & 1.418 & 1.00 & 1.96 & 4659 \\ 
\hline
AP & 200 & tube & 52 & 35 & 200 & 374 & 74 & 0.398 & 7.97E-11 & 1.67E-10 & 21.8 & 6 & 144 & 530 & 1.345 & 1.00 & 1.71 & 2509 \\ 
\hline
AP & 100 & box & 100 & 35 & 100 & 953 & 190 & 0.248 & 1.27E-10 & 5.33E-11 & 7.3 & 3 & 271 & 464 & 1.338 & 1.00 & 1.76 & 325 \\ 
\hline
AP & 100 & box & 500 & 35 & 100 & 4765 & 953 & 0.200 & 1.09E-10 & 3.92E-11 & 41.0 & 6 & 178 & 513 & 1.490 & 1.00 & 1.97 & 743 \\ 
\hline
AP & 100 & box & 50 & 35 & 100 & 476 & 95 & 0.238 & 7.32E-11 & 4.87E-11 & 5.1 & 3 & 108 & 919 & 1.368 & 1.00 & 1.61 & 270 \\ 
\hline
AP & 100 & box & 50 & 20 & 100 & 272 & 54 & 0.179 & 8.94E-11 & 2.96E-11 & 2.6 & 6 & 102 & 505 & 1.489 & 1.00 & 1.91 & 1315 \\ 
\hline
AP & 100 & box & 100 & 35 & 100 & 953 & 190 & 0.249 & 1.89E-10 & 6.35E-11 & 46.0 & 5 & 301 & 1633 & 1.282 & 1.00 & 1.59 & 565 \\ 
\hline
AP & 100 & box & 100 & 35 & 100 & 953 & 190 & 0.251 & 1.13E-10 & 5.77E-11 & 14.2 & 5 & 283 & 740 & 1.546 & 1.00 & 1.95 & 731 \\ 
\hline
AP & 100 & box & 250 & 35 & 100 & 2382 & 476 & 0.348 & 4.12E-10 & 1.60E-10 & 84.5 & 1 & 184 & 184 & 1.454 & 1.45 & 1.45 & 398 \\ 
\hline
AP & 100 & box & 50 & 35 & 100 & 476 & 95 & 0.245 & 1.08E-10 & 5.56E-11 & 9.0 & 2 & 108 & 159 & 1.473 & 1.39 & 1.56 & 237 \\ 
\hline
AP & 20 & box & 50 & 35 & 100 & 476 & 47 & 0.276 & 6.06E-11 & 3.01E-10 & 125.8 & 4 & 2155 & 2580 & 0.765 & 0.72 & 0.86 & 4782 \\ 
\hline
AP & 20 & box & 50 & 35 & 100 & 476 & 95 & 0.157 & 5.81E-11 & 1.86E-11 & 1.6 & 2 & 141 & 1155 & 1.615 & 1.59 & 1.64 & 210 \\ 
\hline
AP & 20 & box & 50 & 35 & 100 & 476 & 95 & 0.178 & 5.81E-11 & 2.57E-11 & 3.8 & 5 & 130 & 838 & 1.338 & 1.00 & 1.59 & 215 \\ 
\hline
AP & 20 & box & 50 & 35 & 100 & 476 & 95 & 0.179 & 6.17E-11 & 2.61E-11 & 1.2 & 4 & 120 & 1176 & 1.454 & 1.00 & 1.64 & 204 \\ 
\hline
AP & 20 & box & 50 & 35 & 100 & 476 & 95 & 0.201 & 7.20E-11 & 5.14E-11 & 44.7 & 7 & 457 & 1860 & 1.336 & 1.00 & 1.91 & 1512 \\ 
\hline
AP & 20 & box & 50 & 35 & 100 & 476 & 95 & 0.207 & 7.91E-11 & 4.00E-11 & 15.2 & 4 & 107 & 4085 & 1.646 & 1.40 & 1.92 & 365 \\ 
\hline
AP & 20 & box & 50 & 20 & 100 & 272 & 54 & 0.134 & 8.78E-11 & 1.35E-11 & 1.8 & 0 &  &  &  &  &  &  \\ 
\hline
AP & 20 & box & 100 & 35 & 100 & 953 & 190 & 0.216 & 8.99E-11 & 6.47E-11 & 155.9 & 5 & 109 & 4567 & 1.588 & 1.30 & 1.94 & 573 \\ 
\hline
AP & 20 & box & 250 & 35 & 100 & 2382 & 476 & 0.221 & 1.29E-10 & 6.15E-11 & 111.5 & 3 & 166 & 8197 & 1.580 & 1.51 & 1.70 & 345 \\ 
\hline
AP & 20 & box & 500 & 35 & 100 & 4765 & 953 & 0.133 & 5.92E-11 & 1.91E-11 & 107.3 & 3 & 229 & 10127 & 1.527 & 1.28 & 1.75 & 223 \\ 
\hline
AP & 20 & box & 50 & 35 & 100 & 476 & 95 & 0.177 & 8.17E-11 & 3.34E-11 & 39.1 & 5 & 387 & 1197 & 1.254 & 1.00 & 1.62 & 1312 \\ 
\hline
AP & 20 & box & 50 & 20 & 100 & 272 & 54 & 0.139 & 4.11E-11 & 1.69E-11 & 1.5 & 0 &  &  &  &  &  &  \\ 
\hline
AP & 20 & tube & 52 & 35 & 200 & 374 & 74 & 0.182 & 6.11E-11 & 2.74E-11 & 4.5 & 4 & 102 & 544 & 1.358 & 1.00 & 1.73 & 1093 \\ 
\hline
NAP & 20 & box & 50 & 35 & 100 & 476 & 95 & 0.278 & 1.18E-10 & 8.74E-11 & 42.3 & 2 & 7090 & 7933 & 1.736 & 1.72 & 1.75 & 363 \\ 
\hline
NAP & 20 & box & 100 & 35 & 100 & 953 & 190 & 0.234 & 9.82E-11 & 6.22E-11 & 70.8 & 4 & 756 & 4044 & 1.472 & 1.23 & 1.77 & 558 \\ 
\hline
NAP & 20 & box & 250 & 35 & 100 & 2382 & 476 & 0.221 & 1.37E-10 & 6.36E-11 & 295.1 & 3 & 1731 & 8573 & 1.568 & 1.29 & 1.84 & 375 \\ 
\hline
NAP & 20 & box & 500 & 35 & 100 & 4765 & 953 & 0.165 & 1.09E-10 & 5.05E-11 & 862.8 & 1 & 3867 & 3867 & 1.177 & 1.18 & 1.18 & 464 \\ 
\hline
NAP & 20 & box & 50 & 35 & 100 & 476 & 95 & 0.176 & 5.70E-11 & 2.72E-11 & 4.7 & 4 & 110 & 1216 & 1.480 & 1.00 & 1.76 & 279 \\ 
\hline
NAP & 20 & box & 50 & 35 & 100 & 476 & 95 & 0.209 & 1.12E-10 & 4.67E-11 & 16.1 & 2 & 1030 & 2223 & 1.708 & 1.59 & 1.83 & 256 \\ 
\hline
NAP & 20 & box & 50 & 35 & 100 & 370 & 74 & 0.345 & 3.06E-10 & 1.46E-10 & 21.0 & 0 &  &  &  &  &  &  \\ 
\hline
NAP & 20 & box & 50 & 35 & 400 & 476 & 95 & 0.398 & 1.72E-10 & 1.50E-10 & 28.8 & 8 & 454 & 1118 & 1.490 & 1.00 & 1.98 & 2180 \\ 
\hline
NAP & 20 & box & 50 & 35 & 400 & 476 & 95 & 0.460 & 1.61E-10 & 2.07E-10 & 15.2 & 3 & 115 & 2879 & 1.552 & 1.30 & 1.85 & 1342 \\ 
\hline
NAP & 20 & box & 50 & 35 & 400 & 370 & 74 & 0.559 & 2.73E-10 & 2.89E-10 & 10.5 & 2 & 1180 & 1567 & 1.753 & 1.66 & 1.85 & 839 \\ 
\hline
NAP & 20 & box & 50 & 35 & 100 & 476 & 95 & 0.267 & 7.93E-11 & 7.36E-11 & 28.3 & 8 & 102 & 3119 & 1.467 & 1.00 & 1.94 & 1759 \\ 
\hline
NAP & 20 & box & 50 & 35 & 100 & 476 & 95 & 0.231 & 1.02E-10 & 4.89E-11 & 21.9 & 4 & 498 & 2012 & 1.562 & 1.28 & 1.88 & 432 \\ 
\hline
NAP & 20 & box & 50 & 35 & 100 & 370 & 74 & 0.268 & 1.46E-10 & 7.13E-11 & 14.0 & 2 & 869 & 1901 & 1.719 & 1.68 & 1.76 & 393 \\ 
\hline
NAP & 20 & box & 50 & 35 & 400 & 476 & 95 & 0.439 & 1.62E-10 & 1.73E-10 & 9.7 & 0 &  &  &  &  &  &  \\ 
\hline
NAP & 20 & box & 50 & 35 & 400 & 476 & 95 & 0.364 & 1.44E-10 & 1.09E-10 & 2.7 & 3 & 189 & 1145 & 1.333 & 1.00 & 1.52 & 536 \\ 
\hline
NAP & 20 & box & 50 & 35 & 400 & 370 & 74 & 0.536 & 2.06E-10 & 2.47E-10 & 10.3 & 2 & 776 & 1308 & 1.757 & 1.69 & 1.83 & 619 \\ 
\hline
NAP & 20 & box & 100 & 35 & 100 & 953 & 190 & 0.304 & 1.91E-10 & 9.95E-11 & 36.1 & 2 & 9172 & 9529 & 1.516 & 1.50 & 1.53 & 471 \\ 
\hline
NAP & 20 & box & 500 & 35 & 100 & 4765 & 953 & 0.225 & 1.44E-10 & 5.95E-11 & 65.6 & 2 & 110 & 9762 & 1.820 & 1.72 & 1.92 & 301 \\ 
\hline
NAP & 20 & box & 50 & 35 & 100 & 476 & 95 & 0.219 & 1.25E-10 & 4.84E-11 & 14.7 & 6 & 105 & 1707 & 1.597 & 1.37 & 1.88 & 513 \\ 
\hline
NAP & 20 & tube & 52 & 35 & 200 & 374 & 74 & 0.204 & 5.31E-11 & 3.09E-11 & 5.3 & 3 & 108 & 474 & 1.770 & 1.55 & 1.91 & 268 \\ 
\hline
NAP & 20 & tube & 52 & 35 & 200 & 374 & 74 & 0.225 & 4.39E-11 & 4.48E-11 & 7.1 & 7 & 123 & 298 & 1.423 & 1.00 & 1.91 & 4797 \\ 
\hline
NAP & 20 & box & 50 & 35 & 100 & 476 & 95 & 0.332 & 1.36E-10 & 1.11E-10 & 34.1 & 5 & 1044 & 2692 & 1.546 & 1.22 & 1.88 & 718 \\ 
\hline
NAP & 20 & box & 50 & 35 & 100 & 476 & 95 & 0.252 & 1.06E-10 & 5.32E-11 & 3.1 & 4 & 113 & 1053 & 1.438 & 1.00 & 1.73 & 296 \\ 
\hline
NAP & 20 & box & 50 & 35 & 100 & 476 & 95 & 0.214 & 6.32E-11 & 4.07E-11 & 12.9 & 4 & 147 & 738 & 1.313 & 1.00 & 1.47 & 360 \\ 
\hline
NAP & 20 & box & 50 & 35 & 100 & 476 & 95 & 0.199 & 7.55E-11 & 3.65E-11 & 20.0 & 5 & 107 & 1049 & 1.325 & 1.00 & 1.73 & 1386 \\ 
\hline
NAP & 20 & box & 50 & 35 & 100 & 476 & 95 & 0.328 & 1.26E-10 & 1.17E-10 & 38.2 & 2 & 11308 & 12913 & 1.879 & 1.84 & 1.92 & 421 \\ 
\hline
NAP & 20 & box & 50 & 35 & 100 & 476 & 95 & 0.278 & 1.18E-10 & 9.06E-11 & 42.3 & 2 & 7090 & 7933 & 1.736 & 1.72 & 1.75 & 378 \\
\hline
\end{tabular}
}
\end{table}
\restoregeometry
\clearpage
}

\clearpage
\printbibliography

@article{Kotsalos2019,
author = {Kotsalos, Christos and Latt, Jonas and Chopard, Bastien},
issn = {10902716},
journal = {Journal of Computational Physics},
month = {dec},
pages = {108905},
publisher = {Academic Press Inc.},
title = {{Bridging the computational gap between mesoscopic and continuum modeling of red blood cells for fully resolved blood flow}},
volume = {398},
year = {2019},
doi = {10.1016/j.jcp.2019.108905}
}

@article{Kotsalos2020,
archivePrefix = {arXiv},
arxivId = {1911.03062},
author = {Kotsalos, Christos and Latt, Jonas and Beny, Joel and Chopard, Bastien},
eprint = {1911.03062},
journal = {Interface Focus (accepted for publication)},
month = {},
number = {Computational Biomedicine},
title = {{Digital Blood in Massively Parallel CPU/GPU Systems for the Study of Platelet Transport}},
year = {2020}
}

@article{Boudjeltia2020,
author = {Boudjeltia, Karim Zouaoui and Kostalos, Christos and Ribeiro, Daniel and Rousseau, Alexandre and Lelubre, Christophe and Sartenaer, Olivier and Piagnerelli, Michael and Dohet-Eraly, Jerome and Dubois, Frank and Tasiaux, Nicole and Chopard, Bastien and Meerhaeghe, Alain Van},
journal = {medRxiv},
month = {apr},
pages = {},
publisher = {},
title = {{Spherization of red blood cells and platelets margination in COPD patients.}},
year = {2020},
doi = {10.1101/2020.04.03.20051748}
}

@article{PalabosArticle,
author = {Latt, Jonas and Malaspinas, Orestis and Kontaxakis, Dimitrios and Parmigiani, Andrea and Lagrava, Daniel and Brogi, Federico and Belgacem, Mohamed Ben and Thorimbert, Yann and Leclaire, S{\'{e}}bastien and Li, Sha and Marson, Francesco and Lemus, Jonathan and Kotsalos, Christos and Conradin, Rapha{\"{e}}l and Coreixas, Christophe and Petkantchin, R{\'{e}}my and Raynaud, Franck and Beny, Jo{\"{e}}l and Chopard, Bastien},
issn = {08981221},
journal = {Computers and Mathematics with Applications},
month = {apr},
publisher = {Elsevier Ltd},
title = {{Palabos: Parallel Lattice Boltzmann Solver}},
year = {2020},
doi = {10.1016/j.camwa.2020.03.022}
}

@book{Taleb2020,
archivePrefix = {arXiv},
arxivId = {2001.10488},
author = {Taleb, Nassim Nicholas},
edition = {1},
eprint = {2001.10488},
isbn = {1544508050},
month = {jun},
pages = {446},
publisher = {STEM Academic Press},
title = {{Statistical Consequences of Fat Tails: Real World Preasymptotics, Epistemology, and Applications (Technical Incerto)}},
url = {https://arxiv.org/abs/2001.10488},
volume = {1},
year = {2020}
}

@misc{Clauset2009,
author = {Clauset, Aaron and Shalizi, Cosma Rohilla and Newman, M. E.J.},
booktitle = {SIAM Review},
issn = {00361445},
month = {jun},
number = {4},
pages = {661--703},
title = {{Power-law distributions in empirical data}},
volume = {51},
year = {2009},
doi = {10.1137/070710111}
}

@book{Kleiber2003,
address = {Hoboken, NJ, USA},
author = {Kleiber, Christian and Kotz, Samuel},
booktitle = {Statistical Size Distributions in Economics and Actuarial Sciences},
isbn = {0471150649},
month = {aug},
publisher = {John Wiley {\&} Sons, Inc.},
series = {Wiley Series in Probability and Statistics},
title = {{Statistical Size Distributions in Economics and Actuarial Sciences}},
url = {http://doi.wiley.com/10.1002/0471457175},
year = {2003},
doi = {10.1002/0471457175}
}

@article{Alstott2014,
author = {Alstott, Jeff and Bullmore, Ed and Plenz, Dietmar},
issn = {19326203},
journal = {PLoS ONE},
month = {jan},
number = {1},
pages = {e85777},
publisher = {Public Library of Science},
title = {{Powerlaw: A python package for analysis of heavy-tailed distributions}},
volume = {9},
year = {2014},
doi = {10.1371/journal.pone.0085777}
}

@article{Chopard2017,
author = {Chopard, Bastien and de Sousa, Daniel Ribeiro and L{\"{a}}tt, Jonas and Mountrakis, Lampros and Dubois, Frank and Yourassowsky, Catherine and {Van Antwerpen}, Pierre and Eker, Omer and Vanhamme, Luc and Perez-Morga, David and Courbebaisse, Guy and Lorenz, Eric and Hoekstra, Alfons G. and Boudjeltia, Karim Zouaoui},
issn = {2054-5703},
journal = {Royal Society Open Science},
month = {apr},
number = {4},
pages = {170219},
publisher = {Royal Society},
title = {{A physical description of the adhesion and aggregation of platelets}},
url = {https://royalsocietypublishing.org/doi/10.1098/rsos.170219},
volume = {4},
year = {2017},
doi = {10.1098/rsos.170219},
}

@article{ZydneyColton,
author = {Zydney, Andrew L. and Colton, Clark K.},
issn = {01919059},
journal = {PCH. Physicochemical hydrodynamics},
number = {1},
pages = {77--96},
title = {{Augmented Solute Transport in the Shear Flow of a Concentrated Suspension.}},
volume = {10},
year = {1988}
}

@article{Affeld2013,
author = {Affeld, Klaus and Goubergrits, Leonid and Watanabe, Nobuo and Kertzscher, Ulrich},
issn = {00219290},
journal = {Journal of Biomechanics},
month = {jan},
number = {2},
pages = {430--436},
publisher = {Elsevier},
title = {{Numerical and experimental evaluation of platelet deposition to collagen coated surface at low shear rates}},
volume = {46},
year = {2013},
doi = {10.1016/j.jbiomech.2012.10.030}
}

@article{Reasor2013,
author = {Reasor, Daniel A. and Mehrabadi, Marmar and Ku, David N. and Aidun, Cyrus K.},
issn = {00906964},
journal = {Annals of Biomedical Engineering},
month = {sep},
number = {2},
pages = {238--249},
publisher = {Springer},
title = {{Determination of critical parameters in platelet margination}},
volume = {41},
year = {2013},
doi = {10.1007/s10439-012-0648-7},
}

@article{Kumar2012,
author = {Kumar, Amit and Graham, Michael D.},
issn = {17446848},
journal = {Soft Matter},
month = {nov},
number = {41},
pages = {10536--10548},
publisher = {Royal Society of Chemistry},
title = {{Margination and segregation in confined flows of blood and other multicomponent suspensions}},
volume = {8},
year = {2012},
doi = {10.1039/c2sm25943e},
}

@article{Vahidkhah2014,
author = {Vahidkhah, Koohyar and Diamond, Scott L. and Bagchi, Prosenjit},
issn = {15420086},
journal = {Biophysical Journal},
month = {jun},
number = {11},
pages = {2529--2540},
publisher = {Biophysical Society},
title = {{Platelet dynamics in three-dimensional simulation of whole blood}},
volume = {106},
year = {2014},
doi = {10.1016/j.bpj.2014.04.028},
}

@article{Mehrabadi2016,
author = {Mehrabadi, Marmar and Ku, David N. and Aidun, Cyrus K.},
issn = {24700053},
journal = {Physical Review E},
month = {feb},
number = {2},
pages = {023109},
publisher = {American Physical Society},
title = {{Effects of shear rate, confinement, and particle parameters on margination in blood flow}},
volume = {93},
year = {2016},
doi = {10.1103/PhysRevE.93.023109},
}

@article{Mehrabadi2015,
author = {Mehrabadi, Marmar and Ku, David N. and Aidun, Cyrus K.},
issn = {15739686},
journal = {Annals of Biomedical Engineering},
month = {oct},
number = {6},
pages = {1410--1421},
publisher = {Kluwer Academic Publishers},
title = {{A Continuum Model for Platelet Transport in Flowing Blood Based on Direct Numerical Simulations of Cellular Blood Flow}},
volume = {43},
year = {2015},
doi = {10.1007/s10439-014-1168-4},
}

@article{Crowl2011,
author = {Crowl, L. and Fogelson, A. L.},
issn = {14697645},
journal = {Journal of Fluid Mechanics},
month = {jun},
pages = {348--375},
publisher = {Cambridge University Press},
title = {{Analysis of mechanisms for platelet near-wall excess under arterial blood flow conditions}},
volume = {676},
year = {2011},
doi = {10.1017/jfm.2011.54}
}

@article{Zhao2012,
author = {Zhao, Hong and Shaqfeh, Eric S.G. and Narsimhan, Vivek},
issn = {10706631},
journal = {Physics of Fluids},
month = {jan},
number = {1},
pages = {011902},
publisher = {American Institute of Physics Inc.},
title = {{Shear-induced particle migration and margination in a cellular suspension}},
url = {http://aip.scitation.org/doi/10.1063/1.3677935},
volume = {24},
year = {2012},
doi = {10.1063/1.3677935}
}

@article{Zhao2011,
author = {Zhao, Hong and Shaqfeh, Eric S.G.},
issn = {15393755},
journal = {Physical Review E - Statistical, Nonlinear, and Soft Matter Physics},
month = {jun},
number = {6},
pages = {061924},
publisher = {American Physical Society},
title = {{Shear-induced platelet margination in a microchannel}},
volume = {83},
year = {2011},
doi = {10.1103/PhysRevE.83.061924}
}

@article{Hampton1966,
author = {Hampton, J. R. and Mitchell, J. R.A.},
issn = {00280836},
journal = {Nature},
number = {5040},
pages = {1000--1002},
publisher = {Nature Publishing Group},
title = {{Modification of the electrokinetic response of blood platelets to aggregating agents}},
volume = {210},
year = {1966},
doi = {10.1038/2101000a0},
}

@article{Jy1995,
author = {Jy, W. and Horstman, L. L. and Homolak, D. and Ahn, Y. S.},
issn = {09537104},
journal = {Platelets},
number = {6},
pages = {354--358},
publisher = {Informa Healthcare},
title = {{Original article: Electrophoretic properties of platelets from normal, thrombotic and ITP patients by doppler electrophoretic light scattering analysis}},
volume = {6},
year = {1995},
doi = {10.3109/09537109509078471},
}

@misc{Betts1968,
author = {Betts, J. J. and Betts, J. P. and Nicholson, J. T.},
booktitle = {Nature},
issn = {00280836},
number = {5160},
pages = {1280--1282},
publisher = {Nature Publishing Group},
title = {{Significance of ADP, plasma and platelet concentration in platelet electrophoretic studies}},
volume = {219},
year = {1968},
doi = {10.1038/2191280b0}
}

@article{Eckstein1991,
author = {Eckstein, E. C. and Belgacem, F.},
issn = {00063495},
journal = {Biophysical Journal},
month = {jul},
number = {1},
pages = {53--69},
publisher = {Cell Press},
title = {{Model of platelet transport in flowing blood with drift and diffusion terms}},
volume = {60},
year = {1991},
doi = {10.1016/S0006-3495(91)82030-6}
}

@article{Koltai2017,
author = {Koltai, Katalin and Kesmarky, Gabor and Feher, Gergely and Tibold, Antal and Toth, Kalman},
issn = {1422-0067},
journal = {International Journal of Molecular Sciences},
month = {aug},
number = {8},
pages = {1803},
publisher = {MDPI AG},
title = {{Platelet Aggregometry Testing: Molecular Mechanisms, Techniques and Clinical Implications}},
url = {http://www.mdpi.com/1422-0067/18/8/1803},
volume = {18},
year = {2017},
doi = {10.3390/ijms18081803}
}

@article{Picker2011,
author = {Picker, Susanne M.},
issn = {14730502},
journal = {Transfusion and Apheresis Science},
month = {jun},
number = {3},
pages = {305--319},
publisher = {Elsevier},
title = {{In-vitro assessment of platelet function}},
volume = {44},
year = {2011},
doi = {10.1016/j.transci.2011.03.006}
}

@article{Breet2010,
author = {Breet, Nicoline J. and {Van Werkum}, Jochem W. and Bouman, Heleen J. and Kelder, Johannes C. and Ruven, Henk J.T. and Bal, Egbert T. and Deneer, Vera H. and Harmsze, Ankie M. and {Van Der Heyden}, Jan A.S. and Rensing, Benno J.W.M. and Suttorp, Maarten J. and Hackeng, Christian M. and {Ten Berg}, Jurri{\"{e}}n M.},
issn = {00987484},
journal = {JAMA - Journal of the American Medical Association},
month = {feb},
number = {8},
pages = {754--762},
publisher = {American Medical Association},
title = {{Comparison of platelet function tests in predicting clinical outcome in patients undergoing coronary stent implantation}},
volume = {303},
year = {2010},
doi = {10.1001/jama.2010.181}
}

@article{Harrison2005,
author = {Harrison, Paul},
issn = {0268960X},
journal = {Blood Reviews},
month = {mar},
number = {2},
pages = {111--123},
publisher = {Churchill Livingstone},
title = {{Platelet function analysis}},
volume = {19},
year = {2005},
doi = {10.1016/j.blre.2004.05.002}
}

@article{Vahidkhah2015,
author = {Vahidkhah, Koohyar and Bagchi, Prosenjit},
issn = {17446848},
journal = {Soft Matter},
month = {mar},
number = {11},
pages = {2097--2109},
publisher = {Royal Society of Chemistry},
title = {{Microparticle shape effects on margination, near-wall dynamics and adhesion in a three-dimensional simulation of red blood cell suspension}},
volume = {11},
year = {2015},
doi = {10.1039/c4sm02686a}
}

@article{Gross2014,
author = {Gross, Markus and Kr{\"{u}}ger, Timm and Varnik, Fathollah},
issn = {0295-5075},
journal = {EPL (Europhysics Letters)},
month = {jan},
number = {6},
pages = {68006},
publisher = {IOP Publishing},
title = {{Fluctuations and diffusion in sheared athermal suspensions of deformable particles}},
volume = {108},
year = {2014},
doi = {10.1209/0295-5075/108/68006}
}

@article{Lee2013,
author = {Lee, Tae Rin and Choi, Myunghwan and Kopacz, Adrian M. and Yun, Seok Hyun and Liu, Wing Kam and Decuzzi, Paolo},
issn = {20452322},
journal = {Scientific Reports},
month = {jun},
number = {1},
pages = {1--8},
publisher = {Nature Publishing Group},
title = {{On the near-wall accumulation of injectable particles in the microcirculation: Smaller is not better}},
volume = {3},
year = {2013},
doi = {10.1038/srep02079}
}

@article{Yeo2010,
author = {Yeo, Kyongmin and Maxey, Martin R.},
issn = {0295-5075},
journal = {EPL (Europhysics Letters)},
month = {nov},
number = {2},
pages = {24008},
publisher = {IOP Publishing},
title = {{Anomalous diffusion of wall-bounded non-colloidal suspensions in a steady shear flow}},
volume = {92},
year = {2010},
doi = {10.1209/0295-5075/92/24008}
}

@article{Dutta2018,
author = {Dutta, Ritabrata and Chopard, Bastien and L{\"{a}}tt, Jonas and Dubois, Frank and {Zouaoui Boudjeltia}, Karim and Mira, Antonietta},
issn = {1664-042X},
journal = {Frontiers in Physiology},
month = {aug},
number = {AUG},
pages = {1128},
publisher = {Frontiers Media S.A.},
title = {{Parameter Estimation of Platelets Deposition: Approximate Bayesian Computation With High Performance Computing}},
url = {https://www.frontiersin.org/article/10.3389/fphys.2018.01128/full},
volume = {9},
year = {2018},
doi = {10.3389/fphys.2018.01128},
}

@article{Weisel2019,
author = {Weisel, J. W. and Litvinov, R. I.},
title = {Red blood cells: the forgotten player in hemostasis and thrombosis},
journal = {Journal of Thrombosis and Haemostasis},
volume = {17},
number = {2},
pages = {271-282},
year = {2019},
doi = {10.1111/jth.14360}
}

@article{Flamm2012,
author = {Flamm, M. H. and Diamond, S. L.},
issn = {15739686},
journal = {Annals of Biomedical Engineering},
month = {mar},
number = {11},
pages = {2355--2364},
pmid = {22460075},
publisher = {Kluwer Academic Publishers},
title = {{Multiscale systems biology and physics of thrombosis under flow}},
volume = {40},
year = {2012},
doi = {10.1007/s10439-012-0557-9}
}

\clearpage
{\small
\section*{\small Acknowledgments}
We acknowledge support from the Swiss National Supercomputing Center
(CSCS, Piz-Daint supercomputer), the National Supercomputing Center in
the Netherlands (Surfsara, Cartesius supercomputer), and the HPC
Facilities of the University of Geneva (Baobab cluster). 

\section*{\small Source Code}
The source code for the statistical analysis is available in GitHub: \url{https://github.com/kotsaloscv/PLTs-FatTails}. The tools for performing the cellular blood flow simulations, as presented in Kotsalos et al. \cite{Kotsalos2019,Kotsalos2020}, are available as part of Palabos open-source library (npFEM specialized module): \url{https://palabos.unige.ch}.

\section*{\small Funding}
This project has received funding from the European Union’s Horizon 2020 research and innovation programme under grant agreement No 823712 (CompBioMed2 project). Furthermore, this work was supported by grants from the CHU Charleroi; the Fonds pour la Chirurgie Cardiaque; Fonds de la Recherche Medicale en Hainaut (FRMH) and by
the Swiss PASC project ``Virtual Physiological Blood: an HPC framework for
blood flow simulations in vasculature and in medical devices''.

\section*{\small Authors' contributions} C.K and B.C. wrote the paper. 
C.K. performed the research,
carried out the simulations and the analysis of the results. 
K.Z.B. supervised the research and revised the
manuscript. R.D. supervised the research and revised the
manuscript. J.L. wrote part of the computational framework,
supervised the research and revised the manuscript. B.C. conceived
and supervised the research and revised the manuscript. All authors
approved the final version of the manuscript.

\section*{\small Competing interests}
The authors declare that they have no competing interests.
}

\end{document}